\documentclass[aps,prl,groupedaddress,twocolumn,10pt,reprint,superscriptaddress, longbibliography,floatfix]{revtex4-2}
\usepackage{epsfig,dsfont,amssymb,amsmath,amsthm,amsfonts,amsbsy,mathrsfs}
\usepackage{amsmath,amssymb,amsthm,mathtools,bm}
\usepackage{lipsum}
\usepackage{color}
\usepackage{graphicx}
\usepackage{amsmath}
\usepackage{amssymb}
\usepackage{subfigure}
\usepackage{float}
\usepackage{hyperref}
\usepackage{verbatim}
\usepackage{soul}
\usepackage{CJK}
\setstcolor{red}

 \usepackage{gensymb}
 \usepackage{comment}
\def\d{{\rm d}}

\renewcommand{\eqref}[1]{\textrm{Eq.}(\ref{#1})}

\newcommand{\R}{\mathbb{R}}
\newcommand{\dd}{\mathrm{d}}
\newcommand{\bb}{\bm}
\newcommand{\transpose}{\mathsf{T}}

\begin{document}
\title{Thermodynamic Limits on Reliable Signaling by Biochemical Traveling Waves}

\author{Shengyao Luo}
\affiliation{
Department of Physics, Tsinghua University, Beijing, 100084, China 
}
\affiliation{
Peking-Tsinghua Center for Life Science, Peking University, Beijing, 100871, China 
}

\author{Yuping Chen}
\affiliation{State Key Laboratory of Quantitative Synthetic Biology, Shenzhen Institute of Synthetic Biology, Shenzhen Institutes of Advanced Technology, Chinese Academy of Sciences, Shenzhen, 518055, China
}

\author{Yuansheng Cao}
\affiliation{
Department of Physics, Tsinghua University, Beijing, 100084, China 
}
\begin{abstract} 
Biochemical traveling waves transmit signals across cells and tissues, but the thermodynamic cost of reliable propagation remains unclear. We develop a stochastic thermodynamic framework for reaction--diffusion systems with stable traveling waves and show that diffusion of the wave position is bounded by the dissipation specifically associated with propagation. The bound follows by projecting noisy field dynamics onto the adjoint translational mode, which maps the wave position to an effective biased random walk. Its tightness is controlled by the non-self-adjoint part of the linearized dynamics, with finite wave speed and antisymmetric reaction dynamics generically producing deviations from equality. For excitable trigger waves in a FitzHugh--Nagumo model, we show that the slow inhibitor can dominate the propagation cost, yielding a trade-off among wave speed, inhibitor amplitude, and dissipation. We test these predictions in stochastic simulations of a microscopic Belousov--Zhabotinsky reaction--diffusion system and find consistent signatures in mitotic trigger-wave experiments in \textit{Xenopus} egg extracts. The same relation further imposes an annihilation-limited bound on the reliable signaling rate of wave trains.
\end{abstract}

\maketitle
\section{Introduction}
Traveling waves are widely used to transmit signals over long distances in cells and tissues.  Examples include action potentials in neurons, mitotic trigger waves in early embryos\cite{chang2013,brantley2021,gelens2014,huang2024robust}, and calcium or protein activity waves that coordinate intracellular and intercellular processes\cite{cooke1976,aulehla2008,oates2012,vergassola2018,di2022waves,leybaert2012,kuga2011,hino2020erk,bolado2020,whitaker2006,liao2016}. Although these systems differ in molecular details, they share a common dynamical feature: they propagate with an approximately stable shape and speed, enabling signals to travel over distances much larger than those accessible by simple diffusion. Such propagation, however, is sustained by biochemical reactions operating far from equilibrium and therefore requires continuous energy consumption.

Many biological traveling waves occur in mesoscopic systems, where stochasticity in the underlying reactions can generate substantial fluctuations. These stochasticity can cause wave-position fluctuations, which reduces the precision of timing and spatial coordination when waves trigger downstream events at reproducible positions or times. For well-mixed nonequilibrium systems, it is well known that fluctuations are constrained by energetic cost through relations such as thermodynamic uncertainty relations\cite{barato2015,cao2015free,gingrich2016,horowitz2017,avanzini2019,horowitz2020,marsland2019}. Whether and how an analogous constraint governs the fidelity of spatially extended traveling waves remains unclear. A key difficulty is that the relevant fluctuating object is not a scalar current, but a collective coordinate: the stochastic position of a propagating field profile.

Here we develop a thermodynamic framework for stochastic biochemical traveling waves. By projecting the noisy dynamics onto the adjoint translational mode, we obtain an effective biased random walk for the wave position and derive an uncertainty relation between the wave-position diffusion constant and the propagation dissipation. This construction also identifies when the bound is tight, indicating that finite wave speed and antisymmetric reaction dynamics generically open a gap. We then specialize the theory to excitable trigger waves, where the slow recovery field controls the refractory tail and can dominate the energetic cost of propagation. Finally, we test the resulting predictions in a reversible Belousov--Zhabotinsky (BZ) reaction-diffusion model and in mitotic trigger-wave data from \textit{Xenopus} egg extracts, and use the same bound to estimate the maximum reliable signaling rate of wave trains limited by wave--wave annihilation.

\section{Analytical results}
\begin{figure}
    \centering
    \includegraphics[width=0.45\textwidth]{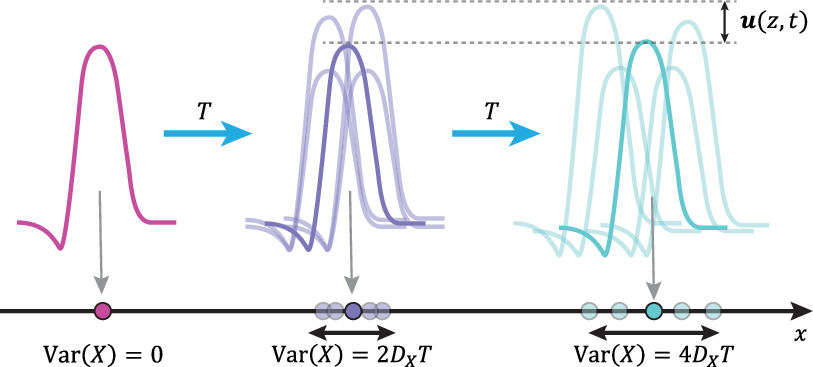} 
    \caption{Fluctuations of a traveling wave can be decomposed into a translational mode, described by the wave-position displacement $X(t)$ in the comoving frame $z=x-ct$, and shape fluctuations $\bm u(z,t)$. The translational mode exhibits diffusive dynamics, $\mathrm{Var}(X)=2D_Xt$, whereas the shape fluctuations remain bounded.}
    \label{fig:mode-decomposition}
\end{figure}
We consider an $n$-component field $\bb a(x,t)\in\R^n$ evolving in one dimension $x\in(-\infty,\infty)$ according to a stochastic partial differential equation
\begin{equation}
\partial_t \bb a \;=\; \bb F(\bb a)\;+\; \bb D\,\partial_x^2\bb a\;+\;\bb \xi(x,t),
\label{eq:spde-generic}
\end{equation}
where $\bb F(\bb a)$ is the reaction dynamics between components and $\bb D$ is the diffusion matrix. We summarize all the reaction stochasticity as the white noise field  $\bb \xi$ with $ \mathbb{E}[\bb\xi_{\alpha}(x,t) \bb\xi_{\beta}(x',t')] = 2\Delta_{\alpha}\delta_{\alpha\beta}\delta(x-x')\delta(t{-}t')$, where $\Delta_{\alpha}$ is the noise strength of component $\alpha$. We assume the deterministic dynamics admit a stable traveling wave $\bar{\bb a}(z)$ with speed $c>0$, where $z=x-ct$ is the comoving coordinate. 

In the weak noise regime, fluctuations of the wave can be separated into stochastic displacement of the wave position (or phase) and deformations of its shape (Fig.~\ref{fig:mode-decomposition}):
\[
    \bm a(x,t)=\bar{\bm a}(z-X(t))+\bm u(z,t),
\]
where $X(t)$ is the relative displacement in the comoving coordinate and $\bm u(z,t)$ represents the shape fluctuations around the steady shape profile.

Linearizing Eq.\ref{eq:spde-generic} around $\bar{\bb a}$ in the comoving coordinate gives the operator $\mathcal L \equiv \left.d \bb F/ d \bb a\right|_{\bar{\bb a}} + \bb D\,\partial_z^2 + c\,\partial_z$. Translational invariance gives the Goldstone mode, $\partial_z \bar{\bb a}(z)$, whose corresponding adjoint mode $\bm \psi(z)$ satisfies $\mathcal L^\dagger\bm\psi=0$ and $\langle{\bb\psi},\partial_z{\bar{\bb a}}\rangle =1$, where $\mathcal L^\dagger$ is the adjoint operator of $\mathcal L$. Because the traveling wave is stable, all non-Goldstone modes remain bounded under weak noise. We use the convention $\langle \bb u, \bb v \rangle \equiv \int_{-\infty}^{\infty} \bb u(z)^{\transpose} \bb v(z) \, \dd z$ throughout this letter. Projecting the stochastic dynamics onto the adjoint mode and integrating out these bounded shape-related modes yields the effective phase equation (Fig.~\ref{fig:mode-decomposition}; details see SI):
\begin{equation}
    \frac{\d X(t)}{\d t}= -\langle\bb \psi, \bb\xi(x,t)\rangle.
\label{eq:Xdot}
\end{equation}
Thus $X(t)$ follows diffusive dynamics with diffusion constant $D_X=\langle \bm\psi,\bm\Delta\bm\psi\rangle$. Equivalently, the laboratory-frame position $Y(t)=ct+X(t)$ behaves as a biased random walk with drift velocity $c$ and diffusion constant $D_X$.

Sustained wave propagation requires nonequilibrium driving from energy dissipation. To quantify this cost, let $P[\bm a, t]$ be the probability distribution of the field. Its dynamics obey the functional Fokker-Planck equation:
\[
\frac{\partial P[\bb a,t]}{\partial t}=-\int \d x \frac{\bb \delta J}{\delta \bb a(x)},
\]
with probability flux
\[
    \bb J(\bb a)=[\bb F(\bb a)+\bb D\partial_x^2\bb a]P[\bb a,t]-\bb\Delta\frac{\delta P[\bb a,t]}{\delta\bb a(x)}.
\]
The total dissipation rate (or entropy production rate) at steady state is then:
\[
    \dot{W}=\int\dd z\int\mathcal{D}\bb a\,\frac{\bb J_{\text{ss}}(\bb a)^{\transpose}\bb\Delta^{-1}\bb J_{\text{ss}}(\bb a)}{P_{\rm ss}(\bb a)}.
\]

The same collective-coordinate decomposition also separates the dissipation into two contributions to leading order in the weak-noise limit: $\dot{W}=\dot{W}_{\text{int}}+\dot{W}_{\text{prop}}$. Here $\dot{W}_{\text{int}}$  denotes the internal dissipation associated with local shape fluctuations of the wave profile. This contribution scales with system size. By contrast, $\dot{W}_{\text{prop}}$ is the dissipation specifically required to translate the stable wave profile forward and is independent of system size (see SI). In the weak-noise limit, this propagation dissipation rate is:
\begin{equation}
    \dot{W}_{\text{prop}}\approx c^2\langle \partial_z \bar{\bb a},\bb \Delta^{-1} \partial_z \bar{\bb a}\rangle.
    \label{eq:w-prop}
\end{equation}

Combining this expression with the wave-position diffusion constant $D_X=\langle \bm\psi,\bm\Delta\bm\psi\rangle$, and using the normalization $\langle{\bb\psi},\partial_z{\bar{\bb a}}\rangle =1$, the Cauchy–Schwarz inequality gives the central thermodynamic bound for traveling waves:
\begin{equation}
    \frac{D_X\dot{W}_{\text{prop}}}{c^2}\geq 1,
    \label{eq:TUR}
\end{equation}
where a generalized form also holds for spatially correlated noise (see SI). \eqref{eq:TUR} establishes a direct trade-off between energetic cost and propagation fidelity: at a fixed wave speed, reducing wave-position diffusion requires a minimum dissipation rate for propagation.

The tightness of \eqref{eq:TUR} is controlled by the non-self-adjointness of the linearized operator $\mathcal{L}$. In the SI, we show that the gap from the lower bound is proportional to the norm of $\mathcal{L}_A \partial_z \bar{\bb a}$, where $\mathcal{L}_A=c\partial_z+ \bb \Delta[(\bm\Delta^{-1}\bm F')-(\bm\Delta^{-1} \bm F')^T]/2$ is the antisymmetric part of $\mathcal{L}$ in the noise-weighted metric. This expression identifies two mechanisms that move the system away from saturation: finite translational motion, represented by $c\partial_z$, and antisymmetric reaction dynamics, represented by the antisymmetric part of $\bm\Delta^{-1}\bm F'$. For a one-component system, the reaction term is scalar and therefore has no antisymmetric part, leaving $\mathcal{L}_A=2c\partial_z$. Thus any finite-speed one-component wave with a nontrivial profile is generically away from the bound. This finite-speed contribution is analogous to the thermodynamic uncertainty relation (TUR) gap of a finite-affinity Markov-jump biased random walk, for which equality is reached only in the linear-response limit \cite{barato2015,gingrich2016, horowitz2020}.  As a concrete example, consider a one-component reaction-diffusion system with a bistable relaxation front (such as Allen-Cahn dynamics), \eqref{eq:TUR} expands as $\frac{D_X\dot{W}_{\text{prop}}}{c^2}\approx 1+\beta \left(\frac{c^2}{\omega D}\right)$, where $\omega$ is the characteristic relaxation rate, $D$ is the real-space diffusion constant, and $\beta>0$ is a model-dependent constant (see SI and Fig. S1). Equality is recovered only in the stall limit $c\to 0$, where the front approaches equilibrium. In multi-component systems, nonreciprocal interactions in $\bb F$ provide an additional antisymmetric contribution to $\mathcal{L}_A$ and therefore produce further deviation from saturation\cite{Macieszczak2018}. 

\begin{figure}[htbp]
    \centering
    \includegraphics[width=0.45\textwidth]{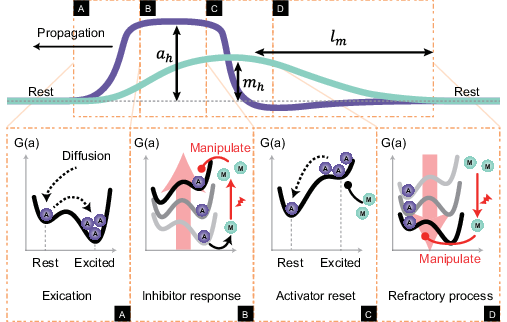} 
    \caption{Schematic of propagation and dissipation in an excitable trigger wave. Top: Activator ($a$, purple) and inhibitor ($m$, green) profiles during propagation, with amplitudes  $a_{\rm h}$ and $m_{\rm h}$.  Bottom: One propagation cycle: diffusion-driven excitation of the activator, inhibitor response, inhibitor-mediated reshaping of the effective landscape $G(a)$, and refractory relaxation. The slow inhibitor field remains elevated over the extended refractory tail behind the waveback (with characteristic length of $l_m$) and dominates the propagation dissipation.}
 \label{fig:fhn}
\end{figure}
While Allen-Cahn dynamics describes a simple relaxation front, many multi-component biological traveling waves are excitable: after passage of the wavefront, the medium must recover to the rest state before it can support another signal. To illustrate the consequences of \eqref{eq:TUR} in this setting, we consider the FitzHugh-Nagumo (FHN) model. The system contains two coupled fields: a fast activator $a$, which drives the wavefront, and a slow inhibitor $m$, which provides delayed negative feedback and resets the medium:
\begin{align}
\label{eq:FHN}
\partial_t a &= D_a\,\partial_x^2 a + \omega_a[f(a)-k m]+\eta_a(x,t),\nonumber \\
\partial_t m &= D_m\,\partial_x^2 m + \omega_m(a-m)+\eta_m(x,t).
\end{align}
Here $f(a)=-a(a-a_c)(a-1)$, $a_c$ is the excitation threshold, $\omega_a,\omega_m$ set the characteristic rate of the activator and inhibitor, and $k$ denotes the feedback strength. We focus on the strong timescale separation, $\varepsilon=\omega_m/\omega_a\ll 1$. This fast-slow structure is biophysically natural for excitable biochemical waves, where a rapid activation step is often followed by slower recovery or inhibition that sets the refractory region. In the strong timescale separation limit, the inhibitor varies slowly during the fast activator transition and acts as a quasi-static control parameter. For fixed $m$, the activator dynamics can be viewed as motion in an effective landscape $G(a)=-\int [f(a)-km]da$. Thus the inhibitor tilts the activator landscape and controls the transition between the rest and excited branches (Fig.~\ref{fig:fhn}).

This separation of timescales also determines where the propagation dissipation is concentrated. The fast activator mainly undergoes a sharp transition between the two branches of the effective landscape, whereas the slow inhibitor remains elevated over the extended inhibitor tail behind the wavefront. As a result, the inhibitor contribution dominates the propagation dissipation (details in the SI, with numerical support in Fig. S2 and Fig. S3). For an approximately exponential inhibitor tail, we obtain (Fig.~\ref{fig:fhn}):
\begin{equation}
    \dot{W}_{\text{prop}}\approx \frac{c\,\omega_m a_{\rm h}m_{\rm h}}{\Delta_m},
    \label{eq:c-peak}
\end{equation}
where $a_{\rm h},m_{\rm h}$ are the peak amplitudes of the activator and inhibitor fields, respectively, and $\Delta_m$ is the inhibitor noise strength. At weak negative feedback (small $k$), the fast activator reaches the
upper excited branch, so that $a_{\rm h}\sim 1$ is saturated and $m_{\rm h}$ is independent of $a_{\rm h}$. At strong 
negative feedback (large $k$), the activator is reset before reaching its saturated value, so that $a_{\rm h}$ is graded and $m_{\rm h}$ is positively correlated with $a_{\rm h}$ (see SI and Fig. S5).


\eqref{eq:c-peak} links the energetic cost of propagation to directly observable wave properties. At fixed propagation dissipation, it predicts a trade-off between wave speed and inhibitor amplitude: faster waves must have a smaller inhibitor amplitude, whereas waves with larger inhibitor amplitude propagate more slowly. This relation also provides a macroscopic estimate of the propagation cost in excitable waves and will be used below to compare theory with simulations and experiments.

\section{Numerical and experimental validation}

\begin{figure}
    \centering
    \includegraphics[width=0.45\textwidth]{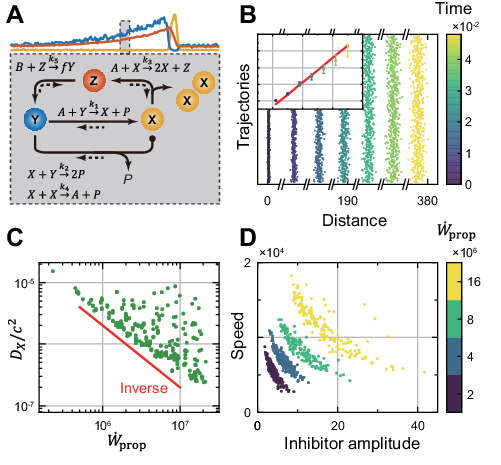} 
    \caption{Numerical validation in the reversible BZ reaction–diffusion model. (A). Reversible BZ model with microscopic reactions. The network is driven by chemostatted species (A and B). Representative spatial profiles of a propagating wave are shown in the top. (B). Peak positions from repeated realizations at different times (color bar); the positional variance grows linearly with time, defining the diffusion constant $D_X$ (inset). (C). Normalized diffusion constant $D_X/c^2$ versus baseline-subtracted propagation dissipation rate $\dot{W}_{\text{prop}}$ across sampled parameter sets, showing the inverse lower bound predicted by \eqref{eq:TUR}. (D). Wave speed versus inhibitor peak amplitude for different $\dot{W}_{\text{prop}}$, consistent with \eqref{eq:c-peak}.}
    \label{fig:bz}
\end{figure}

Because \eqref{eq:w-prop} is defined at the coarse-grained field level, it captures only the coarse-grained dissipation associated with translating the wave profile. It does not include the full microscopic entropy production of the underlying reaction network, including hidden dissipation from eliminated degrees of freedom and baseline housekeeping dissipation required to maintain the nonequilibrium medium \cite{Esposito2012,Kawaguchi2013,rao2016}. We therefore ask whether the bound in \eqref{eq:TUR} and the trigger-wave scaling in \eqref{eq:c-peak} remain valid in systems closer to microscopic and experimental systems. To this end, we examine two excitable systems: a stochastic reaction–diffusion model of the BZ reaction with explicit microscopic reactions, and mitotic trigger waves in {\it Xenopus} egg extracts.

We first test the theory in a stochastic reaction-diffusion model of the BZ reaction, which supports excitable trigger waves under strong activator-inhibitor timsescale separation similar to the FHN model (Fig.~\ref{fig:bz}A and details in the SI). To quantify dissipation, all microscopic reactions and diffusion steps are rendered reversible, with nonequilibrium driving maintained by chemostatted reservoirs \cite{avanzini2019}. This construction allows us to compute the microscopic entropy-production rate  as $\dot{W}=\sum_i(J_i^+-J_i^-)\ln(J_i^+/J_i^-)$, where $J_i^{\pm}$ are the forward and reverse fluxes of the $i$-th reaction or diffusion step. Because the homogeneous rest state is dissipative, we isolate the propagation component $\dot{W}_{\text{prop}}$ by subtracting the rest-state baseline from $\dot{W}$ (details in SI and Fig. S2 \& S3).

From ensembles of waves initiated under identical conditions, we find that the variance of wave position around its mean drift grows linearly in time, consistent with \eqref{eq:Xdot} (Fig.~\ref{fig:bz} B). This linear growth allows us to extract the wave diffusion constant $D_X$. By contrast, shape fluctuations around the translated wave profile remain bounded (Fig. S4), supporting the decomposition into diffusive translational mode and stable non-translational modes. Across sampled parameter sets, the BZ model exhibits the inverse relation between propagation dissipation $\dot{W}_{\text{prop}}$ and normalized wave-position diffusion constant $D_X/c^2$ (Fig.~\ref{fig:bz}C) predicted by \eqref{eq:TUR}. Here, $D_X/c^2$ measures the uncertainty accumulated over a fixed travel distance, because faster waves spend less time exposed to phase diffusion.

Although the BZ model contains more than two chemical species, its wave dynamics reduce effectively to an activator-inhibitor structure under strong timescale separation (see SI and Fig. S5). The simulations also support the speed-amplitude trade-off predicted by \eqref{eq:c-peak} (Fig.~\ref{fig:bz}D): at fixed $\dot{W}_{\text{prop}}$, the wave speed decreases as the inhibitor peak amplitude increases. In addition, increasing rest-state dissipation strongly suppresses spontaneous misfiring events, indicating a separate energetic cost for stabilizing the quiescent state against noise (Fig. S6).

\begin{figure}
    \centering
    \includegraphics[width=0.45\textwidth]{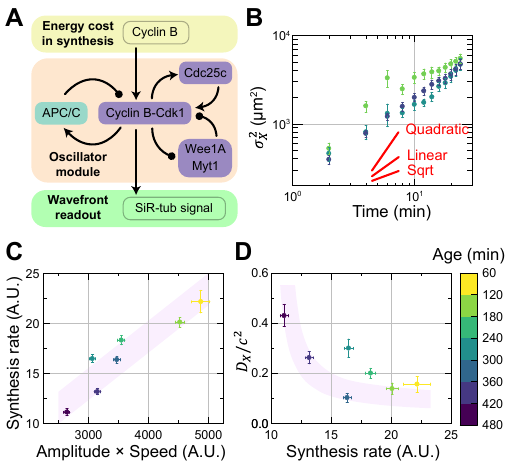} 
    \caption{Experimental test of wave-position diffusion and energetics in mitotic trigger waves in {\it Xenopus} egg extracts. (A). Mitotic trigger waves arise from the coupled feedback network involving Cyclin B-Cdk1, Cdc25, Wee1/Myt1, and APC/C; the propagating wave is read out by the SiR-tubulin signal. (B). Wave positional variance $\sigma_X^2$ grows approximately linearly with propagation time, allowing extraction of a wave-position diffusion constant $D_X$. Data are grouped into different aging stages of the extract (color bar). (C). The protein-synthesis-rate proxy, obtained from the maximal wavefront rising slope, is positively correlated with the speed-amplitude product, consistent with the scaling in \eqref{eq:c-peak}. (D). The normalized diffusion constant $D_X/c^2$ decreases as the inferred energy-supply proxy increases, consistent with \eqref{eq:TUR}. }
    \label{fig:xenopus}
\end{figure}

We next ask whether signatures of the same thermodynamic constraint can be detected in a biological trigger-wave system. We therefore analyze mitotic waves in cycling \textit{Xenopus} egg extracts. In this assay, cytoplasmic extract confined in a long tube supports recurrent waves of mitotic entry and exit, reported by sperm chromatin or polymerized microtubules \cite{huang2024robust}. These waves are generated by the coupled positive and negative feedback loops involving Cyclin B-Cdk1, Cdc25, Wee1, and APC, and propagate approximately linearly through the extract (Fig.~\ref{fig:xenopus}A). This system provides an experimentally accessible realization of biochemical trigger waves. Experimental procedures and data-processing steps are summarized in the SI and illustrated in Fig. S7.

During the experiment, the extract gradually ages over several hours. This aging reduces mitochondrial activity and, consequently, the energy supply available for ATP- and GTP-dependent biochemical processes. We use this slow drift as a natural perturbation by dividing the data into quasi-stationary time windows and comparing wave statistics across different energetic states. Tracking individual waves within each window shows that the positional variance grows approximately linearly with propagation time, allowing us to estimate the wave diffusion constant $D_X$ (Fig.~\ref{fig:xenopus}B; see Fig. S7 and SI).

Because the microscopic dissipation rate cannot be measured directly in the extract, we use empirical proxies for the energetic cost of propagation. First, we use the maximal rising slope of the SiR-tubulin wavefront as an experimentally accessible proxy for the local wavefront activity and the energetic state of the extract. The SiR-tubulin signal reports microtubule polymerization; therefore, its maximal rising slope reflects the rate at which polymerized microtubules accumulate during wave passage. Since microtubule polymerization and turnover are coupled to GTP-consuming tubulin dynamics, this slope provides an indirect proxy for local energetic turnover during propagation. Second, we use \eqref{eq:c-peak} to construct a trigger-wave propagation-cost proxy from the measured wave speed and recovery-wave amplitude. In this assay, high Cyclin B--Cdk1 activity suppresses the SiR-tubulin signal, so the recovery-phase maximum of the SiR-tubulin signal provides an inverse readout of mitotic Cdk1 activity and reports the amplitude of the inhibitor. The observed variation in wave amplitude across time windows suggests that the system is not in a fully activator-saturated regime  (see SI and Fig.~S8); therefore, the inhibitor amplitude remains coupled to the activator excursion. We use the product of wave speed and recovery amplitude as an empirical proxy for the trigger-wave propagation cost. This proxy is positively correlated with the independently defined wavefront-activity proxy across time windows (Fig.~\ref{fig:xenopus}C).

This correlation supports the interpretation that both quantities report the changing energetic state of the propagating wave. Using the trigger-wave propagation-cost proxy, we find that the normalized diffusion constant $D_X/c^2$ decreases as the inferred propagation cost increases (Fig.~\ref{fig:xenopus}D). Thus, in a noisy biological extract, higher energetic activity is associated with reduced wave-position diffusion, consistent with the thermodynamic bound in \eqref{eq:TUR}. 

\section{Thermodynamic limit on reliable wave-based signaling}
Combining \eqref{eq:TUR} and \eqref{eq:c-peak} gives 
\[
\frac{D_X}{c^2}\ge \frac{\Delta_m}{c\omega_m a_{\rm h}m_{\rm h}}
\]
which links the normalized fluctuation of a trigger wave to its kinetic parameters and inhibitor amplitude. At fixed noise strength, increasing the wave speed or the inhibitor amplitude lowers the minimum uncertainty accumulated over a given propagation distance. Such a constraint is relevant when traveling waves are used to trigger events with reproducible timing or position, including cardiac excitation waves \cite{davidenko1992,pertsov1993}, mitotic waves in embryogenesis\cite{chang2013,di2022waves}, and patterning waves in somitogenesis\cite{cooke1976,oates2012,liao2016}. 

Wave-position diffusion has an additional consequence in systems that transmit signals through repeated trigger waves. When information is encoded in the timing or frequency of a wave train, positional fluctuations can change the spacing between neighboring waves. Because trigger waves are followed by a refractory region of elevated inhibitor field over a length $l_m$ (Fig. 2), a trailing wave that enters this refractory wake is extinguished. Such wave–wave annihilation removes pulses from the train and thereby degrades signaling fidelity.

For reliable transmission over a distance $L$, the spacing between two successive waves $\Delta x$ must exceed the sum of the refractory length $l_m$ and the typical positional diffusion accumulated over the propagation time $L/c$. This requirement imposes an upper bound on the firing frequency $f=c/\Delta x$. Using \eqref{eq:TUR} and \eqref{eq:c-peak} into this spatial constraint, we obtain the annihilation-limited firing rate:
\begin{equation}  
    f_{\max}=\frac{c}{\sqrt{2LD_X/c}+l_m}\leq\frac{c}{\sqrt{2Lc/\dot{W}_{\text{prop}}}+l_m} 
    \label{eq:f-W}
\end{equation}
Thus, increasing propagation dissipation can raise the maximum reliable rate by reducing wave-position diffusion, but only until the refractory length becomes the dominant limitation.

\eqref{eq:f-W} identifies two limiting regimes. For fast waves with a long refractory length, the maximal rate is controlled mainly by $l_m$, and wave-position diffusion gives only a small correction. This appears to be the case for unmyelinated neuronal action potentials ($c\approx 0.7m/s\; D_X\approx 1.7\mu m^2/s$, $l_m\approx 1.4$mm, and $L\approx 1$mm, data from \cite{radivojevic2017}), for which the estimated $f_{\max}\approx 500$Hz, well above typical physiological firing rates (1-10 Hz). By contrast, for slow intracellular cortical waves ($c\approx0.02\sim 0.2\mu$m/s) that propagate over distances comparable to the cell size ($L\sim 20\mu$m), the positional diffusion term can become comparable to, or even larger than, the refractory constraint\cite{yang2018,gerhardt2014,weiner2007,miao2019}. In this regime, the maximal reliable signaling rate is limited more directly by the energetic cost to suppress wavefront diffusion.

\section{Discussion}
In this work, we derived a thermodynamic relation between the positional diffusion of biochemical traveling waves and the dissipation required for propagation. By projecting stochastic reaction–diffusion dynamics onto the adjoint translational mode, we showed that wave-position diffusion is bounded by a minimal propagation dissipation. Numerical simulations and analysis of biological trigger waves are consistent with this prediction.

Conceptually, our result is closely related to recent efforts to formulate stochastic thermodynamics and thermodynamic uncertainty relations at the level of spatially extended fields\cite{niggemann2020,niggemann2022,nardini2017,markovich2021,liang2024}. Instead of following a scalar current in a well-mixed reaction network or Fourier-mode decomposition\cite{niggemann2020,niggemann2022}, we follow a collective coordinate in a fluctuating field and identify the conjugate dissipation associated with its directed motion. This perspective may be useful more broadly for noisy spatiotemporal structures in active media. In particular, the same construction should extend naturally to other traveling patterns, including oscillatory waves, curved fronts, spiral waves, and rotating structures\cite{cross1993,tyson1988}. In those cases, translational and rotational Goldstone modes would lead to coupled collective coordinates, and the corresponding bounds would constrain joint fluctuations of position, phase, or orientation.

The present theory has several limitations. The derivation assumes a stable traveling wave with weak fluctuations, so that shape modes remain bounded and the dynamics can be reduced to an effective random walk for the wave position. Under strong noise, near the loss of excitability, or close to wave instability or front breakup, this reduction may no longer hold\cite{tyson1988,lindner2004,keener2000,yang2003,fenton2002multiple}. In addition, the current formulation is restricted to one spatial dimension. In higher dimensions, front curvature, transverse roughening, and geometry can all affect propagation \cite{keener1986,panja2004,meron1992}. In such systems, one must distinguish more carefully among front diffusion, shape fluctuations, and rotational or orientational motion. Extending the present framework to these cases is a natural direction for future work.

The bound in \eqref{eq:TUR}, and its signaling consequence in \eqref{eq:f-W}, concern the irreducible energetic cost required to suppress wave-position diffusion during propagation. They do not include all energetic costs required for reliable signaling in an excitable medium. Additional dissipation may be needed to maintain the excitable state, regenerate chemical gradients, sustain nonequilibrium fluxes, or suppress spontaneous nucleation and spurious firing. Our numerical results already illustrate this distinction: the cost of preventing spontaneous activation can be separated from the propagation cost itself. More generally, the total implementation cost of biochemical information transmission can substantially exceed the lower bound set by wave-position diffusion. This point is related to recent work showing that the energetic cost of transmitting information through a physical channel can far exceed the Landauer limit, because it is determined by the dissipative physics of the channel rather than by logical irreversibility alone \cite{bryant2023}.

This distinction is especially relevant for the biological validation. In the {\it Xenopus} system, the thermodynamic quantities are not measured directly; instead, the comparison relies on empirical quantities that report the energetic state of the extract and on the trigger-wave propagation-dissipation scaling derived from the theory. A more direct experimental connection between measurable biochemical fluxes, ATP consumption, and propagation dissipation remains an open challenge. Developing such measurements would provide a stronger test of the framework and help determine how closely the inferred propagation cost tracks the underlying nonequilibrium driving in living media.

More broadly, our result places biochemical traveling waves within the thermodynamic landscape of biological information transmission. Over short distances, passive diffusion can transmit molecular signals, with fluctuations governed by equilibrium transport. Over longer distances or for directed cargo transport, molecular motors provide active motion, with precision-dissipation relation given by TURs. At still larger scales, fluid flows introduce macroscopic transport and energetic constraints. Biochemical traveling waves occupy an intermediate regime: they transmit information collectively over mesoscopic distances, and their reliability is limited by stochastic diffusion that can be suppressed by free energy cost. Understanding how living systems combine diffusion, motor-driven transport, fluid flow, and wave-based signaling across scales remains an important open question.


\section{Acknowledgments}
We thank Shiling Liang and Yuhai Tu for helpful comments. S. L., Y. Chen, and Y. Cao are supported by the National Key Research and Development Program of China (Grant No.2024YFA0919600).

\bibliography{wave_ref}

\end{document}


\title{Supporting Information for ``Thermodynamic Limits on Reliable Signaling by Biochemical Traveling Waves"}
\maketitle

\section{Field method for traveling wave system}

We consider an $n$-component continuous field $\bb a(x,t) \in \R^n$ whose spatiotemporal evolution is governed by the following stochastic partial differential equation (SPDE) in one dimension:
\begin{equation}
    \partial_t \bb a \;=\; \bb F(\bb a) \;+\; \bb D\,\partial_x^2\bb a \;+\; \bb \xi(x,t), \quad x\in\R, \ t\ge 0,
    \label{eq:spde_general}
\end{equation}
where $\bb F(\bb a)$ represents the deterministic nonlinear reaction kinetics, $\bb D$ is the diffusion matrix, and $\bb \xi(x,t)$ denotes a Gaussian spatiotemporal noise. All projections below use the standard $L^2$ inner product over the comoving coordinate $z$:
\begin{equation}
    \langle \bb u, \bb v \rangle \equiv \int_{-\infty}^{\infty} \bb u(z)^{\transpose} \bb v(z) \, \dd z.
    \label{eq:standard_inner_product}
\end{equation}

In the deterministic limit, we assume that the system supports a stable traveling-wave profile $\bar{\bb a}(z)$ with speed $c>0$, where $z=x-ct$. The profile satisfies:
\begin{equation}
    \bb F(\bar{\bb a}) \;+\; \bb D\partial_z^2\bar{\bb a} \;+\; c\partial_z \bar{\bb a} = \bb 0.
\end{equation}
Linearizing the comoving dynamics around $\bar{\bb a}(z)$ defines
\begin{equation}
    \mathcal L \equiv \bb D\,\partial_z^2 + c\partial_z + \bb F'(\bar{\bb a}),
\end{equation}
where $\bb F'(\bar{\bb a})$ is the Jacobian of the reaction kinetics. Translational invariance gives the right zero mode, or Goldstone mode,
\begin{equation}
    \mathcal{L}\partial_z\bar{\bb a} = \bb 0.
\end{equation}

The corresponding adjoint zero mode $\bb\psi(z)$ selects the translational component of a noisy perturbation. It satisfies $\mathcal{L}^\dagger\bb\psi=\bb0$, where $\mathcal{L}^\dagger=D\,\partial_z^2 - c\partial_z + \bb F'(\bar{\bb a})$ donates the adjoint operator, and is normalized as
\begin{equation}
    \langle \bb\psi, \partial_z \bar{\bb a} \rangle = \int_{-\infty}^{\infty} \bb \psi(z)^{\transpose} \partial_z \bar{\bb a}(z) \, \dd z = 1.
\end{equation}

In the presence of weak noise, we separate the deterministic translation, the stochastic phase displacement $X(t)$, and the bounded shape perturbation $\bm u$ by writing
\begin{equation}
\bm a(x,t) = \bar{\bm a}(z - X(t)) + \bm u(z,t).
\end{equation}
Here $X(t)$ is the stochastic displacement relative to the uniformly translating frame, so the laboratory-frame wave position is $Y(t)=ct+X(t)$. To uniquely specify the phase coordinate $X(t)$ and eliminate the purely translational zero mode from the shape perturbation $\bb u(z,t)$, we impose the gauge condition $\langle \boldsymbol{\psi}, \bm u \rangle = 0$. Substituting this decomposition into the comoving SPDE and expanding to linear order in $\bm u$ gives
\begin{equation}
\left[\partial_t - \mathcal{L}\right]\bm u = \dot{X} \partial_z \bar{\bm a} + \boldsymbol{\xi}(z+ct, t).
\label{eq:linearized_spde}
\end{equation}

Projecting this equation onto $\boldsymbol{\psi}$ removes the stable shape modes and isolates the stochastic phase dynamics. The gauge condition gives $\partial_t \langle \boldsymbol{\psi}, \bm u \rangle = 0$, while $\langle \boldsymbol{\psi}, \mathcal{L}\bm u \rangle = \langle \mathcal{L}^\dagger \boldsymbol{\psi}, \bm u \rangle = 0$. Using $\langle \boldsymbol{\psi}, \partial_z \bar{\bm a} \rangle = 1$, we obtain
\begin{equation}
\dot{X}(t) = -\langle \boldsymbol{\psi}, \boldsymbol{\xi}(z+ct, t) \rangle = - \int_{-\infty}^{\infty} \boldsymbol{\psi}(z)^\transpose \boldsymbol{\xi}(z+ct,t) \, \mathrm{d}z.
\label{eq:Xdot}
\end{equation}
Equivalently, the full wave position obeys $\dot{Y}(t)=c+\dot{X}(t)$.

If the inherent noise is assumed to be spatiotemporal white noise with a local covariance
\begin{equation}
    \mathbb{E}[\bb\xi(x,t) \bb\xi(x',t')^{\transpose}] = 2\bb\Delta \delta(x-x')\delta(t-t'),
\end{equation}
where $\bb\Delta = \mathrm{diag}(\Delta_1, \dots, \Delta_n)$ is the noise intensity matrix, then $\dot{X}(t)$ is a temporal white-noise process. The same diffusion constant describes the wandering of $Y(t)$ around its mean drift $ct$. The long-time variance defines the wave-position diffusion constant $D_X$:
\begin{equation}
    \langle [X(t)-X(0)]^2 \rangle = 2 D_X t, \quad \text{for large } t.
\end{equation}
The projection formula gives
\begin{equation}
    D_X = \int_{-\infty}^{\infty} \bb\psi(z)^{\transpose} \bb \Delta \bb\psi(z) \, \dd z \equiv \langle \bb \psi, \bb \Delta \bb \psi \rangle.
    \label{eq:DX}
\end{equation}
Thus wave-position diffusion is controlled by how the local noise metric $\bb \Delta$ projects onto the adjoint translational mode, not simply by the noise amplitude of the most visible field component.

\section{Thermodynamic formulation of propagation dissipation}

The projection framework also identifies which part of the entropy production is conjugate to wave translation. With $k_BT=1$, we use ``dissipation rate'' and ``entropy production rate'' interchangeably. Assuming the noise intensity matrix is spatially homogeneous, denoted by the diagonal matrix $\bb \Delta = \mathrm{diag}(\Delta_\alpha)$, the total thermodynamic dissipation rate $\dot{W}$ is given by
\begin{equation}
    \dot{W} = \int_{-\infty}^{+\infty} \mathrm{d}z \int \mathcal{D}\bb a \, \frac{\bb J_{ss}(\bb a)^\transpose \bb \Delta^{-1} \bb J_{ss}(\bb a)}{P_{ss}(\bb a)}.
    \label{eq:total_dissipation}
\end{equation}
Here, $P_{ss}(\bb a)$ is the steady-state functional probability distribution, and $\bb J_{ss}(\bb a)$ represents the steady-state functional probability flux. For a traveling wave, the useful separation is between fluxes that translate the mean profile and fluxes that circulate or relax within the comoving profile. Because the deterministic wave propagates at a constant velocity $c$, we write the Fokker-Planck equation in the comoving coordinate frame $z = x - ct$:
\begin{equation}
    \frac{\partial P}{\partial t} = - \int \mathrm{d}z \left[\frac{\delta }{\delta \bb a(z)} \cdot \bb J^{\text{int}}(\bb a(z))\right].
    \label{eq:field_fpk}
\end{equation}
In this frame, $\bb J^{\text{int}}$ denotes the internal reaction-diffusion flux inside the comoving wave profile, whereas the profile translation contributes a convective flux
\begin{equation}
    \bb J^{\text{prop}} = c\,(\partial_z \bb a)\,P(\bb a).
\end{equation}
The laboratory-frame steady flux can therefore be written as $\bb J=\bb J^{\text{int}}-\bb J^{\text{prop}}$, with the internal flux in the comoving frame given by
\begin{equation}
    \bb J^{\text{int}} = [\bb F(\bb a) + c\partial_z \bb a + \bb D \partial^2_z \bb a] P(\bb a) - \bb \Delta \frac{\delta P(\bb a)}{\delta \bb a}.
\end{equation}

To obtain the functional probability distribution, we decompose the shape fluctuation field $\bb u(z,t)$ into a complete basis of eigenfunctions of the linearized operator $\mathcal{L}$:
\begin{equation}
    \bb u(z,t) = \sum_k A_k(t) \bb u_k(z), \quad \text{with } A_k(t) = \langle \bb u_k^\dagger, \bb u(\cdot, t) \rangle.
\end{equation}
The eigenvalues are assumed to be negative, $\lambda_k < 0$ (for $k > 0$), representing the convergent relaxation of the wave shape. The modes are bi-orthogonalized such that $\langle \bb u_j^\dagger, \bb u_k \rangle = \delta_{jk}$. Under this expansion, applying the functional chain rule $\frac{\delta}{\delta \bb a(z)} = -\bb \psi(z) \frac{\partial}{\partial X} + \sum_k \bb u_k^\dagger(z) \frac{\partial}{\partial A_k}$ transforms the infinite-dimensional field Fokker-Planck equation into a multi-dimensional mode Fokker-Planck equation governing the stochastic amplitudes $A_k(t)$:
\begin{equation}
    \frac{\partial P(A_0,A_1,A_2,\cdots)}{\partial t} = - \sum_{k} \frac{\partial}{\partial A_k} (\lambda_k A_k P) + \sum_{j,k} Q_{jk} \frac{\partial^2 P}{\partial A_j \partial A_k}.
    \label{eq:mode_FP}
\end{equation}
Here, the noise covariance is projected onto the adjoint basis, defining the matrix elements $Q_{jk} = \langle \bb u_j^\dagger, \bb \Delta \bb u_k^\dagger \rangle$. The Goldstone mode ($k=0$) corresponds to the pure spatial translation $\partial_z \bar{\bb a}$, where $\lambda_0 = 0$, $A_0(t) = -X(t)$, and the bare effective spatial diffusion is exactly given by $Q_{00} = D_X$. When extracting the steady-state distribution $P_{ss}$ for $t \to \infty$, the purely diffusive translation mode $X(t)$ decouples from the structural stationary profile. Consequently, the steady state is exclusively determined by the stable shape modes ($k > 0$). The resulting probability distribution is Gaussian, $P_{ss}(\mathbf{A}) \propto \exp \left( -\frac{1}{2}\sum_{jk} A_j (\bb \Sigma^{-1})_{jk} A_k \right)$, where the covariance matrix $\bb \Sigma_{jk} = \langle A_j A_k \rangle$ balances the projected noise $\mathbf{Q}$ against the deterministic damping rates $\bb \Lambda = \mathrm{diag}(\lambda_1, \lambda_2, \dots)$ via the continuous Lyapunov equation:
\begin{equation}
    \bb \Lambda \bb \Sigma + \bb \Sigma \bb \Lambda^\transpose = -2\mathbf{Q} \implies \bb \Sigma_{jk} = - \frac{2Q_{jk}}{\lambda_j + \lambda_k}.
    \label{eq:lyapunov}
\end{equation}
This Ornstein-Uhlenbeck construction will be used to characterize internal dissipation in the stable shape.

To formulate the thermodynamic dissipation, we substitute the probability flux decomposition $\bb J_{ss} = \bb J^{\text{int}}_{ss} - \bb J^{\text{prop}}_{ss}$ into the generalized dissipation rate. This partitions the total entropy production rate into an internal dissipation rate, a cross term, and the propagation dissipation rate:
\begin{equation}
\begin{aligned}
    \dot{W} &= \int \mathrm{d}z \int \mathcal{D}\bb a \, \frac{(\bb J^{\text{int}}_{ss} - \bb J^{\text{prop}}_{ss})^\transpose \bb \Delta^{-1} (\bb J^{\text{int}}_{ss} - \bb J^{\text{prop}}_{ss})}{P_{ss}} \\
    &= \dot{W}_{\text{int}} + \dot{W}_{\text{cross}} + \dot{W}_{\text{prop}}.
\end{aligned}
\end{equation}

The internal dissipation rate $\dot{W}_{\text{int}}$ is the cost of irreversible circulation within the comoving shape degrees of freedom. Utilizing the mode Fokker-Planck equation, the internal probability flux in stable-mode space is
\begin{equation}
    \bb J^{\text{int}}_{\text{mode}} = (\bb \Lambda + \mathbf{Q} \bb \Sigma^{-1}) \mathbf{A} P_{ss}
    \equiv \mathcal{V}_{\text{mode}}\mathbf{A}P_{ss}.
\end{equation}
Here $\mathbf{A}$ is the vector of stable shape amplitudes. The operator $\mathcal{V}_{\text{mode}}$ generates a steady probability current; its circulation is the mode-space signature of broken detailed balance. Under a diagonalizable linearized operator and within the biorthogonal Ornstein-Uhlenbeck approximation, this gives
\begin{equation}
    \dot{W}_{\text{int}} = \int \mathrm{d}\mathbf{A}\, \frac{(\bb J^{\text{int}}_{\text{mode}})^\transpose \mathbf{Q}^{-1} \bb J^{\text{int}}_{\text{mode}}}{P_{ss}} = -\frac{1}{2} \sum_{i,j} \frac{(\lambda_i - \lambda_j)^2}{\lambda_i + \lambda_j} (\mathbf{Q}^{-1})_{ij} Q_{ji}.
\end{equation}
The indices run over stable shape modes only, and the minus sign reflects the convention $\lambda_k<0$. The resulting dissipation is non-negative. This term should be read as the internal entropy production rate of the coarse-grained SPDE. It can remain positive even when the macroscopic wave is absent and the profile no longer translates, because the spatially homogeneous rest state can itself be a nonequilibrium steady state.

The internal dissipation is extensive in the system size. In the spatially homogeneous rest state, the eigenfunctions of $\mathcal{L}$ are plane waves labeled by a continuous wavenumber $q = 2\pi n / L$ and a chemical species index $\alpha$, so that $\lambda_i=\lambda_\alpha(q)$ and $Q_{ij}=\delta_{q,q'}Q_{\alpha\beta}(q)$. In the large-system limit, $\sum_q \approx (L/2\pi)\int \mathrm{d}q$, which gives
\begin{equation}
    \dot{W}_{\text{int}} = L \left[ -\frac{1}{4\pi} \int \mathrm{d}q \sum_{\alpha, \beta} \frac{(\lambda_\alpha(q) - \lambda_\beta(q))^2}{\lambda_\alpha(q) + \lambda_\beta(q)} (\mathbf{Q}^{-1}(q))_{\alpha\beta} Q_{\beta\alpha}(q) \right]
 = L\dot{w}_{\text{int}}.
\end{equation}
Here $\dot{W}_{\text{int}}$ is a total internal dissipation rate, whereas $\dot{w}_{\text{int}}$ is the corresponding internal dissipation rate density. The density $\dot{w}_{\text{int}}$ depends on the intrinsic properties of the chemical network and is independent of the system size. Therefore, the total internal dissipation rate is extensive in the no-wave regime. If $\dot{w}_{\text{int}}>0$, $\dot{W}_{\text{int}}=L\dot{w}_{\text{int}}$ diverges in the thermodynamic limit as $L\to\infty$. This divergence is the continuum signature of a nonequilibrium background rather than a cost specific to wave propagation.

The density $\dot{w}_{\text{int}}$ is closely related to the housekeeping heat density of the underlying reaction network, but it should not be identified with the full microscopic housekeeping heat. The SPDE is a diffusion approximation that keeps the resolved Gaussian fluctuations and the resolved probability currents. Hidden reaction cycles and discrete jump statistics are not represented explicitly. Thus $\dot{w}_{\text{int}}$ should be interpreted as a coarse-grained estimate of the housekeeping dissipation density. The full microscopic housekeeping heat is generally expected to be larger when those hidden channels are restored.

By contrast, the propagation dissipation is the cost of translating the stable wave profile through the thermodynamic noise metric. At leading order in weak noise, the convective flux is evaluated on the mean profile, $\bb J^{\text{prop}}_{ss} = c (\partial_z \bar{\bb a}) P_{ss}$. Substitution into the definition yields:
\begin{equation}
    \dot{W}_{\text{prop}} = c^2 \int_{-\infty}^{\infty} \mathrm{d}z \int \mathcal{D}\bb a \, P_{ss}(\bb a) (\partial_z \bar{\bb a})^\transpose \bb \Delta^{-1} (\partial_z \bar{\bb a}) = c^2 \int_{-\infty}^{\infty} (\partial_z \bar{\bb a})^\transpose \bb \Delta^{-1} (\partial_z \bar{\bb a}) \, \mathrm{d}z.
\end{equation}

The cross term is
\begin{equation}
    \dot{W}_{\text{cross}} = -2 \int \mathrm{d}z \int \mathcal{D}\bb a \frac{(\bb J^{\text{int}}_{ss})^\transpose \bb \Delta^{-1} \bb J^{\text{prop}}_{ss}}{P_{ss}} = -2c \int \mathrm{d}z (\partial_z \bar{\bb a})^\transpose \bb \Delta^{-1} \langle \mathcal{V} \bb u \rangle_{ss}.
\end{equation}
Within the centered linear mode approximation, $\langle \bb u \rangle_{ss}=\bb0$ and the flux velocity is linear in $\bb u$, so the cross term vanishes at this order, $\dot{W}_{\text{cross}}=0$. Corrections would come from nonlinear shape-mode statistics rather than from the leading phase-projection bound. The result corresponds to the dynamical decoupling between the neutral translational mode and the stable shape modes.

In the weak-noise limit, $\dot{W}_{\text{int}}$ scales as $\bb \Delta^0$, whereas $\dot{W}_{\text{prop}}$ scales explicitly as $\bb \Delta^{-1}$. The central bound below uses only $\dot{W}_{\text{prop}}$, because this is the component conjugate to directed motion of the wave profile. Internal dissipation can still be large or functionally important, but it does not directly suppress translational diffusion.

\section{Relation between wave-position diffusion and dissipation}

The bound follows most directly by using the same inner-product notation introduced above. For white noise with covariance matrix $\bb\Delta$, define the noise-weighted inner product
\begin{equation}
    \langle \bb u,\bb v\rangle_{\bb\Delta^{-1}}
    \equiv \int_{-\infty}^{\infty}\bb u(z)^\transpose\bb\Delta^{-1}\bb v(z)\,\dd z .
    \label{eq:delta_weighted_inner_product}
\end{equation}
Then the two relevant observables are
\begin{align}
    D_X &= \langle \bb\Delta\bb\psi,\bb\Delta\bb\psi\rangle_{\bb\Delta^{-1}},\\
    \frac{\dot{W}_{\text{prop}}}{c^2}
    &= \langle \partial_z\bar{\bb a},\partial_z\bar{\bb a}\rangle_{\bb\Delta^{-1}}.
\end{align}
The phase normalization is the mixed inner product in the same metric,
\begin{equation}
    \langle \bb\Delta\bb\psi,\partial_z\bar{\bb a}\rangle_{\bb\Delta^{-1}}
    = \langle \bb\psi,\partial_z\bar{\bb a}\rangle = 1.
\end{equation}
Applying the Cauchy-Schwarz inequality in this Hilbert space gives
\begin{equation}
    D_X\frac{\dot{W}_{\text{prop}}}{c^2}
    \ge
    \langle \bb\Delta\bb\psi,\partial_z\bar{\bb a}\rangle_{\bb\Delta^{-1}}^2
    =1.
    \label{eq:TUR_bound}
\end{equation}

The same argument applies to spatially correlated noise. Let $\bb C$ be a symmetric positive-definite covariance operator, and define
\begin{equation}
    \langle \bb u,\bb v\rangle_{\bb C^{-1}}
    \equiv \int\!\!\int \bb u(z)^\transpose\bb C^{-1}(z-z')\bb v(z')\,\dd z\,\dd z' .
\end{equation}
With $(\bb C\bb\psi)(z)=\int \bb C(z-z')\bb\psi(z')\,\dd z'$, the wave-position diffusion constant and dissipation metrics become
\begin{align}
    D_X &= \langle \bb C\bb\psi,\bb C\bb\psi\rangle_{\bb C^{-1}},\\
    \frac{\dot{W}_{\text{prop}}}{c^2}
    &= \langle \partial_z\bar{\bb a},\partial_z\bar{\bb a}\rangle_{\bb C^{-1}}.
    \label{eq:Wprop_correlated}
\end{align}
The normalization remains $\langle \bb C\bb\psi,\partial_z\bar{\bb a}\rangle_{\bb C^{-1}}=1$. Cauchy-Schwarz therefore gives the identical bound, provided the same positive-definite covariance operator defines both wave-position diffusion and propagation dissipation.

\section{Geometric origin of bound non-saturation}

The bound is saturated only when the noise-weighted adjoint mode is parallel to the translational mode, $\bb \Delta \bb \psi \propto \partial_z \bar{\bb a}$. This condition is geometric: it relates the alignment of the mode that collects noise into wave-position diffusion to the mode whose translation costs propagation dissipation.

To quantify the loss of this alignment, we use the noise-weighted inner product defined in \eqref{eq:delta_weighted_inner_product}. The adjoint operator in this metric is $\mathcal{L}^* = \boldsymbol{\Delta} \mathcal{L}^{\dagger} \boldsymbol{\Delta}^{-1}$, and the transformed adjoint mode satisfies $\mathcal{L}^*(\bb\Delta\bb\psi)=0$. We decompose $\mathcal{L}$ into symmetric and anti-symmetric parts in this metric,
\begin{equation}
    \mathcal{L}_S = \frac{1}{2}(\mathcal{L}+\mathcal{L}^*), \qquad
    \mathcal{L}_A = \frac{1}{2}(\mathcal{L}-\mathcal{L}^*).
\end{equation}

Projecting $\boldsymbol{\Delta}\boldsymbol{\psi}$ onto the right translational mode gives
\begin{equation}
    \bb \Delta \bb \psi = \alpha  \partial_z \bar{\bb a} + \bb \Phi_\perp,
\end{equation}
where $\boldsymbol{\Phi}_{\perp}$ is orthogonal to $\partial_z\bar{\bb a}$ under $\langle\cdot,\cdot\rangle_{\boldsymbol{\Delta}^{-1}}$. The biorthogonality normalization gives
\begin{equation}
    \alpha || \partial_z \bar{\bb a} ||^2_{\boldsymbol{\Delta}^{-1}} = 1,
    \qquad
    \alpha = \frac{c^2}{\dot{W}_{\text{prop}}}.
\end{equation}
The macroscopic diffusion coefficient can then be written as
\begin{equation}
    D_X = ||\boldsymbol{\Delta} \boldsymbol{\psi}||^2_{\boldsymbol{\Delta}^{-1}}
    = \alpha^2 ||\partial_z \bar{\bb a}||^2_{\boldsymbol{\Delta}^{-1}} + ||\boldsymbol{\Phi}_{\perp}||^2_{\boldsymbol{\Delta}^{-1}}.
\end{equation}
Multiplying both sides by $\dot{W}_{\text{prop}}/c^2$ yields an exact expression for the gap above the bound:
\begin{equation}
    \frac{D_X \dot{W}_{\text{prop}}}{c^2} = 1 + \frac{1}{\alpha}||\boldsymbol{\Phi}_{\perp}||^2_{\boldsymbol{\Delta}^{-1}}.
\end{equation}
Thus, saturation is equivalent to $\boldsymbol{\Phi}_{\perp}=0$. Any dynamical mechanism that rotates $\bb\Delta\bb\psi$ away from $\partial_z\bar{\bb a}$ increases $D_X$ at fixed $\dot{W}_{\text{prop}}$.

The source of this orthogonal deviation is the anti-symmetric dynamics of the linearized operator in the noise-weighted metric. Since $\mathcal{L}\partial_z\bar{\bb a}=0$ and $\mathcal{L}^*(\bb\Delta\bb\psi)=0$, substituting the projection above gives
\begin{equation}
    \boldsymbol{\Phi}_{\perp} = 2\alpha (\mathcal{L}^*)^{-1}_{\perp} \mathcal{L}_A \partial_z \bar{\bb a}.
\end{equation}
Here $(\mathcal{L}^*)^{-1}_{\perp}$ denotes the pseudoinverse on the nonzero-mode subspace selected by the phase gauge, $\langle \partial_z \bar{\bb a}, \mathcal{L}_A \partial_z \bar{\bb a}\rangle_{\bb\Delta^{-1}}=0$, as the $\mathcal{L}_A$ is an asymmetric operator in this noise-weighted matric. Inserting this solution into the bound-gap identity gives
\begin{equation}
    \frac{D_X \dot{W}_{\text{prop}}}{c^2} - 1 = 4 \frac{  ||(\mathcal{L}^*)^{-1}_{\perp} \mathcal{L}_A \partial_z \bar{\bb a}||^2_{\boldsymbol{\Delta}^{-1}} }{ ||\partial_z \bar{\bb a}||^2_{\boldsymbol{\Delta}^{-1}} }.
\end{equation}
The excess factor above unity is set by how strongly $\mathcal{L}_A$ rotates the adjoint mode away from the translational direction.

For the reaction-diffusion operator used here, and assuming that diffusion is symmetric in the noise-weighted metric, the anti-symmetric operator can be written as
\begin{equation}
    \mathcal{L}_A = \frac{1}{2}(\mathcal{L} - \mathcal{L}^*) = c\partial_z + \frac{1}{2} \left(\bb F' -\bb \Delta (\bb F')^\transpose \bb \Delta^{-1} \right) = c\partial_z + \bb \Delta \hat{\bb F}'_A,
    \qquad
    \hat{\bb F}'_A \equiv \frac{1}{2}\left(\bb\Delta^{-1}\bb F'-(\bb F')^\transpose\bb\Delta^{-1}\right).
\end{equation}
The first term, $c\partial_z$, is present even for a scalar relaxation front and represents the non-self-adjointness generated by finite propagation speed. This is the operator-level expression of propagation-induced spatial asymmetry. The second term, $\bb\Delta\hat{\bb F}'_A$, represents non-reciprocal reaction dynamics in the thermodynamic force metric. Non-reciprocal reaction dynamics can provide an additional field-specific mechanism when $\hat{\bb F}'_A\partial_z\bar{\bb a}$ has a component in the stable subspace. In both cases, the gap grows through mode mis-alignment.

\section{Wave-position diffusion and dissipation in Allen-Cahn model}

The scalar Allen-Cahn front provides a useful solvable case because its adjoint mode and propagation dissipation can be computed explicitly. We consider the noisy Allen-Cahn equation:
\begin{equation}
    \partial_t a = \omega_a f(a) + D_a \partial_x^2 a + \eta_a(x,t).
    \label{eq:NagumoForm}
\end{equation}
Here $f(a)=-a(a-a_c)(a-1)$, and $a_c \in (0,1/2)$ sets the local threshold and the energy bias between the two homogeneous states. This bias gives the front a finite velocity. In the deterministic limit ($\eta_a = 0$), the system supports a stable front propagating at velocity $c$:
\begin{equation}
    \bar{a}(z) = \frac{1}{2}\left[ 1 - \tanh\left(\frac{z}{l}\right)\right].
    \label{eq:WavefrontAnsatz_a}
\end{equation}
Here $l=\sqrt{8D_a/\omega_a}$ is the front width and $c=(1-2a_c)\sqrt{\omega_aD_a/2}$ is set by the threshold bias.

For this scalar front, the forward and adjoint linear operators are
\begin{align}
    \mathcal{L}   &= D_a\,\partial_z^2 + c\,\partial_z + \omega_a\,f'(\bar{a}(z)), \\
    \mathcal{L}^\dagger &= D_a\,\partial_z^2 - c\,\partial_z + \omega_a\,f'(\bar{a}(z)).
    \label{eq:L-Ldag_Nagumo}
\end{align}
The right zero mode is $\partial_z\bar a$. The adjoint zero mode, which filters translational diffusion from shape fluctuations, is
\begin{equation}
    \psi(z) = \frac{1}{\Psi_0}\exp(\gamma z)\partial_z\bar a(z), \qquad \gamma = c/D_a.
\end{equation}
Here the constant $\Psi_0$ is induced to satisfy the normalization $\langle \psi,\partial_z\bar a\rangle=1$. It is easy to check that $\mathcal{L^\dagger} \psi=0$.

At the phase-coexistence point $a_c=1/2$, the front is stationary ($c=0$), $\mathcal L$ is self-adjoint, and the adjoint mode reduces to the direct translational mode.

Assuming white noise with intensity $\Delta_a$ and imposing $\langle \psi,\partial_z\bar a\rangle=1$, the wave-position diffusion constant is
\begin{equation}
    D_X = \Delta_a\, \frac{\displaystyle \int_{-\infty}^{\infty} e^{2\gamma z}\,(\partial_z\bar{a})^2\,\dd z}{\displaystyle \left(\int_{-\infty}^{\infty} e^{\gamma z}\,(\partial_z\bar{a})^2\,\dd z\right)^2} = 96\Delta_a l \frac{\big(1-\gamma^2 l^2\big)\tan\big(\pi\gamma l/2\big)}{\pi\gamma l\big(4-\gamma^2 l^2\big)^2}.
\end{equation}

The propagation dissipation rate is the cost of translating this concentration gradient at velocity $c$:
\begin{equation}
    \dot{W}_{\text{prop}} = \frac{c^2}{\Delta_a}\int_{-\infty}^{\infty}{(\partial_z\bar{a})^2\,\dd z} = \frac{c^2}{3l\Delta_a}.
\end{equation}

Combining the two expressions gives the exact relation between wave-position diffusion and dissipation for this model:
\begin{equation}
    \frac{D_X \dot{W}_{\text{prop}}}{c^2} = 32\frac{\big(1-\gamma^2 l^2\big)\tan\big(\pi\gamma l/2\big)}{\pi\gamma l\big(4-\gamma^2 l^2\big)^2} = \frac{(4a_c - 1)(4a_c - 3) \tan(2\pi a_c)}{16\pi (1 - 2a_c) a_c^2 (1 - a_c)^2}.
\end{equation}
As $a_c\to 1/2$ and therefore $c\to0$, the ratio approaches the equality value $1$. The ratio remains well-defined because both the drift and the propagation cost vanish with $c$, while their scaled product approaches the equilibrium alignment condition. 

To further characterize the near-equilibrium regime, an asymptotic expansion of this exact analytical ratio around the stationary limit ($c \to 0$) yields a parabolic scaling relation:
\begin{equation}
    \frac{D_X \dot{W}_{\text{prop}}}{c^2} \approx 1 + \beta \frac{c^2}{\omega_a D_a},
\end{equation}
where the dimensionless coefficient is positive, $\beta = 2(\pi^2-6)/3 \approx 2.58$. This analytical reduction provides a stringent check of the relation $\frac{D_X \dot{W}_{\text{prop}}}{c^2} \geq 1$. Because the Allen-Cahn model is scalar, $\bb F'$ has no matrix-valued non-reciprocal reaction dynamics. The bound gap is therefore generated solely by the propagation-induced spatial asymmetry associated with finite $c$, consistent with the leading correction above being proportional to $c^2$. The stochastic simulations in Fig.~S1 illustrate the wavefront definition, the diffusive growth of wave-position variance, and the inverse relation between $D_X/c^2$ and propagation dissipation in this scalar setting.

\section{FHN application: time-scale separation and propagation dissipation}

We next use a FitzHugh-Nagumo (FHN) trigger wave to show how time-scale separation controls propagation dissipation. We define a fast activator $a(x,t)$ and a slow inhibitor $m(x,t)$, with $\bb a=(a,m)^\transpose$:
\begin{align}
    \partial_{t} a &= \omega_{a}[f(a)-km] + D_a\partial_{x}^{2} a + \eta_a(x,t), \\
    \partial_{t} m &= \omega_{m}(a-m) + D_m\partial_{x}^{2} m + \eta_m(x,t). \label{eq:FHN_m}
\end{align}
Here $f(a)=-a(a-a_c)(a-1)$, $a_c$ is the excitation threshold, $k$ is the feedback strength, and $\omega_a,\omega_m$ are the activator and inhibitor's characteristic rate. The noise sources are independent with covariance $\mathbb{E}[\eta_\alpha(x_1,t_1)\eta_\beta(x_2,t_2)] = 2\Delta_\alpha \delta_{\alpha\beta}\delta(x_1-x_2)\delta(t_1-t_2)$. Thus $\bb \Delta = \mathrm{diag}(\Delta_a,\Delta_m)$.

For this two-field system, the propagation dissipation rate follows directly from the general formula:
\begin{equation}
    \dot{W}_{\text{prop}} = c^2 \int_{-\infty}^{\infty} (\partial_z \bar{\bb a})^{\transpose} \bb \Delta^{-1} (\partial_z \bar{\bb a}) \, \dd z
    = \int_{-\infty}^{\infty} \left[ \frac{c^2}{\Delta_a}(\partial_z \bar{a})^2 + \frac{c^2}{\Delta_m}(\partial_z \bar{m})^2 \right] \dd z.
    \label{eq:FHN_Wprop}
\end{equation}
The general wave-position diffusion and dissipation relation applies to this FHN wave without further model-specific assumptions. The useful FHN-specific question is how $\dot{W}_{\text{prop}}$ is distributed across the fast activator front and the slow refractory tail.

Excitable trigger waves usually operate with strong time-scale separation, $\omega_a\gg\omega_m$ (example see Fig. S5 as $\varepsilon_1 \ll 1$). The activator changes sharply near the wavefront. The inhibitor relaxes slowly over a long refractory tail whose length is set by $c/\omega_m$. In this regime the profile integral in \eqref{eq:FHN_Wprop} is often dominated by the inhibitor-channel
\begin{equation}
    \dot{W}_{\text{prop}}^{(m)} \equiv \frac{c^2}{\Delta_m} \int_{-\infty}^{\infty} (\partial_z \bar{m})^2 \, \dd z,
    \label{eq:FHN_Wprop_recovery}
\end{equation}
so that $\dot{W}_{\text{prop}}\simeq\dot{W}_{\text{prop}}^{(m)}$ when the slow refractory tail dominates the profile integral. This inhibitor-dominated scaling is reflected in the BZ simulations shown below in Fig.~S2 and Fig.~S3.

We now evaluate \eqref{eq:FHN_Wprop_recovery} analytically within a piecewise profile approximation. We neglect $D_m\partial_z^2\bar m$ in the slow inhibitor equation, hold the activator at $a_{\rm h}$ in the excited interval $0<z<L$, and set the activator to its rest value in the refractory tail $z<0$. The inhibitor equation then reduces to
\begin{align}
    c\partial_z \bar{m} + \omega_m(a_{\rm h} - \bar{m}) &= 0, \quad 0 < z < L, \\
    c\partial_z \bar{m} - \omega_m \bar{m} &= 0, \quad z < 0.
\end{align}
With $\lambda=c/\omega_m$ and boundary conditions $\bar m(-\infty)=0$, $\bar m(0)=m_{\rm h}$, and $\bar m(L)=0$, the corresponding piecewise solution is
\begin{equation}
    \bar{m}(z) =
    \begin{cases}
        a_{\rm h} - (a_{\rm h} - m_{\rm h})\e^{z/\lambda}, & 0 < z < L, \\
        m_{\rm h}\e^{z/\lambda}, & z < 0,
    \end{cases}
    \label{eq:FHN_m_profile}
\end{equation}
where $m_{\rm h}=\bar m(0)$ is the peak inhibitor level at the onset of the refractory tail. The condition $\bar m(L)=0$ fixes $m_{\rm h}=a_{\rm h}(1-\e^{-L/\lambda})$. Equivalently, $\e^{L/\lambda}=a_{\rm h}/(a_{\rm h}-m_{\rm h})$. The piecewise construction enforces continuity of $\bar m$ at $z=0$. The resulting dissipation is therefore a closed-form approximation to the smooth FHN pulse.

Substituting this profile into \eqref{eq:FHN_Wprop_recovery} gives an explicit sum of the excited-plateau and refractory-tail contributions,
\begin{equation}
    \int_{-\infty}^{\infty}(\partial_z\bar m)^2\,\dd z
    =
    \frac{(a_{\rm h}-m_{\rm h})^2}{2\lambda}\big(\e^{2L/\lambda}-1\big)
    + \frac{m_{\rm h}^2}{2\lambda}.
\end{equation}
Therefore, within the piecewise approximation,
\begin{equation}
    \dot{W}_{\text{prop}}\approx \frac{c\,\omega_m}{2\Delta_m} \Big[ (a_{\rm h}-m_{\rm h})^2 \big(\e^{2L/\lambda}-1\big) + m_{\rm h}^2 \Big]= \frac{c\,\omega_m\,a_{\rm h}m_{\rm h}}{\Delta_m}.
    \label{eq:FHN_Wprop_piecewise}
\end{equation}
Equation \eqref{eq:FHN_Wprop_piecewise} shows that the slow inhibitor kinetics and inhibitor amplitude set the leading propagation cost. 

The FHN Jacobian contains directed activator-inhibitor couplings through $-k\omega_a$ and $\omega_m$. In the noise-weighted metric these couplings generically produce non-reciprocal reaction dynamics, $\hat{\bb F}'_A\ne\bb0$. Thus FHN waves provide a natural setting where non-reciprocal activator-inhibitor couplings provide an additional source of non-saturation. As a comparison, the scalar Allen-Cahn isolates the finite-speed contribution to the bound gap (Fig. S1).

\section{BZ reaction and numerical procedure}
The reversible BZ model is used as a controlled numerical system in which both wave diffusion and propagation dissipation can be measured at microscopic level. We consider the classical reduced BZ reaction network, a paradigmatic reaction-diffusion system capable of sustaining excitable traveling waves.

The reduced chemical mechanism contains three dynamic intermediates, $X$ ($\text{HBrO}_2$, the autocatalyst), $Y$ ($\text{Br}^-$, the inhibitor), and $Z$ (the oxidized metal ion catalyst, e.g., $\text{Ce}^{4+}$), together with chemostatted reservoir species $A$ ($\text{BrO}_3^-$), $B$, and $P$ ($\text{HOBr}$):
\begin{equation}
    \begin{array}{rll}
    A+Y &\xrightarrow {k_{1}} & X+P, \\
    X+Y &\xrightarrow{k_{2}} & 2P, \\
    A+X &\xrightarrow{k_{3}} & 2 X+Z, \\
    X+X &\xrightarrow{k_{4}} & A+P, \\
    B+Z &\xrightarrow{k_{5}} & f Y,
    \end{array}
    \label{eq:reaction_BZ}
\end{equation}
where $k_i$ represent the effective reaction rate constants, and $f$ is the stoichiometric factor governing the nonlinear feedback strength from the inhibitor. 

We denote the dimensional concentration field of species $\alpha$ by $C_\alpha(r,t)$ in this section. To classify the dynamical regimes, we nondimensionalize the continuous fields by $X_0 = k_3 A/(2k_4)$, $Y_0 = k_3 A/k_2$, and $Z_0 = (k_3 A)^2/(2k_4 k_5)$. We rescale time by $\tau=t/T_0=k_5t$ and space by $\rho=r\sqrt{k_5/D_{\rm chem,X}}$, where $T_0=1/k_5$ is the nondimensionalized time constant, and $D_{\rm chem,X}$ is the chemical diffusion constant of species $X$ and should not be confused with the wave-position diffusion constant $D_X$. The dimensionless equations are
\begin{align}
    \dfrac{\partial x}{\partial \tau} &= \frac{1}{\varepsilon_1} (x-x^2-xy+qy) + \nabla^2 x, \\
    \dfrac{\partial y}{\partial \tau} &= \frac{1}{\varepsilon_2}(-qy-xy+fz) + d_y \nabla^2 y,\\
    \dfrac{\partial z}{\partial \tau} &= x-z + d_z \nabla^2 z ,
\end{align}
where $\varepsilon_1 = k_5/(k_3 A)$, $\varepsilon_2 = qk_5/(k_1 A)$, and $q=2k_1k_4/(k_2k_3)$ is the dimensionless reaction threshold. In the parameter regime used here, spatial coupling is dominated by the autocatalyst $X$, so we take $d_y\approx d_z\approx0$. The main dynamical controls are then the time-scale parameter $\varepsilon_1$ and the threshold $q$. The corresponding wave-supporting parameter region and activator-inhibitor amplitude relation are summarized in Fig.~S5.

To define chemical affinities and entropy production, we render the reduced reaction network reversible. In practical simulations, the reverse reaction rate coefficients are set to a uniformly small value, $k_{\rho-}\equiv k_- = 0.5$, with units determined by the kinetic order of each reverse reaction.

In the deterministic continuum limit, the free-energy dissipation rate $\dot{W}$ is the spatial integral of a dissipation rate density, with both reaction and diffusion contributions and $k_BT=1$:
\begin{equation}
\begin{aligned}
    \dot{W} &= \int_{-\infty}^{\infty} (\dot{w}_{\text{reac}} + \dot{w}_{\text{diff}}) \, \dd r \\
    &= \int_{-\infty}^{\infty} \left[ \sum_{\rho=1}^5 \Delta J_{\rho} \ln{\frac{J_{\rho+}}{J_{\rho-}}} + \sum_i \frac{D_i}{C_i} \left( \frac{\partial C_i}{\partial r} \right)^2 \right] \dd r \\
    &= \int_{-\infty}^{\infty} \left[ \sum_{\rho=1}^5 \Delta J_{\rho} \left( \ln{\frac{k_{\rho+}}{k_{\rho-}}} - \sum_\alpha \Delta S^{\rho}_\alpha \ln{C_\alpha } \right) - \sum_i D_i \frac{\partial^2 C_i}{\partial r^2} \ln{C_i} \right] \dd r, \label{eq:energy_1}
\end{aligned}
\end{equation}
Here $\rho$ labels reactions, $\alpha$ labels all chemical species, and $i\in\{X,Y,Z\}$ labels the dynamic intermediates. The forward and backward fluxes obey mass-action kinetics, $J_{\rho\pm}=k_{\rho\pm}\prod_\alpha(C_\alpha)^{S^{\rho\pm}_\alpha}$, with net flux $\Delta J_\rho=J_{\rho+}-J_{\rho-}$ and stoichiometric difference $\Delta S^\rho_\alpha=S^{\rho-}_\alpha-S^{\rho+}_\alpha$. The spatial profiles of the species and the separate local reaction and diffusion contributions to the dissipation density are shown in Fig.~S2A,B.

For a stable pulse propagating at constant velocity $c$, the deterministic concentration fields satisfy
\begin{equation}
    D_i \frac{\partial^2 C_i}{\partial r^2} + c \frac{\partial C_i}{\partial r} + \sum_\rho \Delta S^\rho_i \Delta J_\rho = 0. \label{eq:common_wave_form}
\end{equation}
Substituting \eqref{eq:common_wave_form} into \eqref{eq:energy_1} eliminates the second-order spatial derivatives. After integration by parts, the total dissipation rate can be written in terms of the spatially integrated reaction currents:
\begin{equation}
    \dot{W} = \sum_{\rho=1}^5 J^{tot}_{\rho} \ln{\frac{\hat{k}_{\rho+}}{\hat{k}_{\rho-}}} - \sum_i c \Delta(C_i \ln{C_i} - C_i), \label{eq:energy_2}
\end{equation}
Here $J^{tot}_\rho=\int\Delta J_\rho\,\dd r$ is the spatially integrated reaction current, and $\Delta g\equiv g(-\infty)-g(+\infty)$ fixes the sign convention for the boundary term in \eqref{eq:energy_2}. The effective pseudo-first-order rate constants $\hat{k}_{\rho\pm}=k_{\rho\pm}\prod_j(C_j)^{S^{\rho\pm}_j}$ absorb the chemostatted pool species $j\in\{A,B,P\}$. For monostable excitable pulses, the boundary states at $r\to\pm\infty$ coincide, so the boundary term vanishes regardless of this convention.

Integrating \eqref{eq:common_wave_form} over space imposes $\sum_\rho \Delta S_i^\rho J_\rho^{tot}=c\Delta C_i=0$. This constraint leaves two independent thermodynamic cycle currents, so the total dissipation rate can be expressed as
\begin{equation}
    \begin{aligned}
        \dot{W} &= (J^{tot}_1 \ln{\Gamma_1} + J^{tot}_5 \ln{\Gamma_2}) \\
        &\approx \left[ \hat{k}_1 Y_{tot} \ln{\frac{\hat{k}_1 k_4 \hat{k}_{2r}}{\hat{k}_{1r} \hat{k}_{4r} \hat{k}_2}} + \hat{k}_5 Z_{tot} \ln{\frac{\hat{k}_3 \hat{k}_5 k_2^f k_4^{(1-f)/2}}{k_{3r} k_{5r} \hat{k}_{2r}^f \hat{k}_{4r}^{(1-f)/2}}} \right], \label{eq:energy_3}
    \end{aligned}
\end{equation}
where the final line uses the strong-driving approximation $J_\rho^{tot}\approx J_{\rho+}$. Here $Y_{tot}=\int C_Y(r)\,\dd r$ and $Z_{tot}=\int C_Z(r)\,\dd r$ are integrated particle numbers.
Because the background rest state of the BZ reaction is itself a nonequilibrium steady state, the spatial integral over a large domain contains an extensive background contribution. Following the general decomposition above, we write the large-domain rate schematically as $\dot{W}=L_{\rm sys}\dot{w}_{\rm rest}+\dot{W}_{\text{prop}}$, where $\dot{w}_{\rm rest}$ is the rest-state internal dissipation rate density and the localized excess is the propagation dissipation rate. In simulations, $\dot{W}_{\text{prop}}$ is extracted operationally by baseline subtraction, replacing $Y_{tot}$ by $Y_{tot}'=\int [C_Y(r)-C_{Y,\rm rest}]\,\dd r$ and $Z_{tot}$ by $Z_{tot}'=\int [C_Z(r)-C_{Z,\rm rest}]\,\dd r$. This subtraction is visualized by the rest-state baseline and the spatially integrated excess propagation dissipation in Fig.~S2C,D.

For typical BZ regimes used here, the affinity of the second cycle dominates, $\Gamma_2\gg\Gamma_1$. The propagation dissipation is therefore mainly governed by the slow $Z$-dependent inhibitor loop. Under the adiabatic tail approximation, $C_Z(r)-C_{Z,\rm rest}\approx Z_{\rm peak}\exp(\hat{k}_5r/c)$ behind a wavefront located at $r=0$ with $r<0$ in the tail, where $Z_{\rm peak}$ denotes the excess peak above the rest state (Fig.~S2D). Hence $Z_{tot}'\propto cZ_{\rm peak}/\hat{k}_5$, and consequently $\dot{W}_{\text{prop}}\propto cZ_{\rm peak}$ up to fixed kinetic and affinity factors. This parallels the FHN result that the slow refractory tail carries the dominant propagation cost (Fig.~S3).

For stochastic validation, we simulate the reaction-diffusion dynamics with a spatial Gillespie algorithm. The one-dimensional domain is discretized into compartments of length $\Delta x$. Chemical reactions within each compartment are evaluated using volume-scaled mesoscopic propensities, and diffusion is represented by nearest-neighbor hopping with rate $D_\alpha/(\Delta x)^2$ for species $\alpha$. The temporal evolution is computed using the exact Gillespie Direct Method.

To remove initial transients, we first integrate the deterministic continuum equations until a stable pulse forms, then initialize the stochastic simulations from this profile. During propagation we track the macroscopic front coordinate $X(t)$, defined as the localized center of mass of the active pulse, together with the species profiles. These observables give the wave-position diffusion constant $D_X$ and the baseline-subtracted propagation dissipation rate $\dot{W}_{\text{prop}}$, allowing a direct numerical test of the relation between the two. The same simulations show that non-translational amplitude fluctuations remain bounded while the wave position diffuses (Fig.~S4).

Furthermore, we test how the residue dissipation in the no-wave regime determines the rest-state stabilization. Although the main bound involves the propagation dissipation of a traveling wave, the no-wave regime also carries the rest-state internal dissipation density, $\dot{w}_{\rm rest}$. In simulations, we find that this residue dissipation can suppress rare fluctuations that cross the excitation threshold and nucleate spontaneous pulses (Fig.~S6A). To quantify the rest-state stabilization effect, we measure noise-induced spontaneous firing by initializing systems near the homogeneous excitable rest state. To make firing statistics comparable across parameter sets, we reduce the reaction length scale to increase the number of observable events and count spontaneous firings per unit length and unit time, $f_{\rm fire}$. We then nondimensionalize this rate using the activator relaxation time $\tau_a=T_0\varepsilon_1$ and the corresponding activation length $\ell_c=(D_{\rm chem,X}T_0\varepsilon_1 q)^{1/2}$, where $T_0, \varepsilon_1, q$ are nondimensionalized constants defined before. The same local time-scale and diffusion-length construction parallels theoretical works in the FHN application above. Plotting relative spontaneous firing probability, $f_{\rm fire}\tau_a$, against the  relative unit dissipation at the rest-state, $\dot{w}_{\rm rest}\tau_a\ell_c$, shows an approximately exponential decrease of the effective firing frequency with increasing rest-state dissipation (Fig.~S6B). Thus, in addition to paying an energy cost for reliable propagation, the nonequilibrium medium can spend energy in the resting state to stabilize the excitable background against noise-driven misfiring.

\section{Mitotic trigger wave and experimental procedure}
\subsection{Xenopus egg extract preparation}

All animal experiments adhered to national and international guidelines under Stanford University Administrative Panel on Laboratory Animal Care protocol 13307. \textit{Xenopus laevis} adults ($>$3 years old; NASCO females LM00531 and males LM00715) were used for this study. Demembranated sperm chromatin was prepared as described previously and typically used at $\sim$100 sperm/$\mu$L extract. \textit{Xenopus} egg extracts were prepared according to established protocols with specific modifications. In brief, freshly laid eggs were dejellied in 20 mg/mL L-cysteine (pH 7.8) for less than 5 min and immediately washed three times in 0.2x MMR solution (20 mM NaCl, 400 $\mu$M KCl, 400 $\mu$M CaCl\textsubscript{2}, 200 $\mu$M MgCl\textsubscript{2}, 1 mM HEPES, 20 $\mu$M EDTA, pH 7.8). Dejellied eggs were activated with 0.5 $\mu$g/mL calcium ionophore A23187 (Sigma C7522) in 0.2x MMR solution for 2 min. Following activation, eggs were washed twice in extract buffer (50 mM sucrose, 100 mM KCl, 100 $\mu$M CaCl\textsubscript{2}, 1 mM MgCl\textsubscript{2}, 10 mM HEPES-KOH, pH 7.7). Eggs were packed via low-speed centrifugation (200x g for 1 min, followed by 600x g for 30 s), and excess buffer was removed. Crushing was performed at 16,000x g at 4 $^\circ$C for 15 min. The cytoplasmic fraction was carefully collected and supplemented with protease inhibitors (leupeptin, pepstatin, chymostatin) and cytochalasin B at 10 $\mu$g/mL each. Finally, extracts were clarified by centrifugation (16,000x g at 4 $^\circ$C for 5 min) once or twice prior to experimentation.

\subsection{Visualization and tracking of mitotic wavefront}

Cycling extracts were used to monitor mitotic trigger waves. Microtubule networks were visualized using 200 nM SiR-tubulin (CY-SC002, Cytoskeleton, Inc). The dissolution of these networks at the onset of mitosis provided a sensitive readout for mitotic wavefronts. PTFE tubes (100 $\mu$m inner diameter) were filled with the extract, immersed in heavy mineral oil within imaging chambers, and imaged at room temperature. Biological replicates were defined as extracts prepared from distinct egg clutches obtained from different female frogs. Kymographs of propagating trigger waves were generated from time-lapse videos using the Multi Kymograph tool in FIJI/ImageJ and then analyzed with a custom Python pipeline (Fig.~S7A). Because the exact frame interval and microscope pixel calibration for the processed kymographs were not fixed independently, the experimental analysis below is reported in frame and processed-pixel units.

The kymographs were normalized, smoothed along the spatial direction, and scanned for local intensity extrema. Wavefront positions were identified as half-maximum points on falling transitions from local peaks to following troughs (Fig.~S7B). The same extrema provided the wave amplitude $A$ and the maximal rising slope. Candidate points were filtered for spatiotemporal continuity, refined by Gaussian kernel density estimation, and linked across adjacent frames by nearest-neighbor matching. Tracks shorter than 15 frames were excluded from the quantitative analysis.

In the notebook implementation, time is measured in frames and space is measured in processed kymograph pixels. Because the code samples the spatial axis as \texttt{kymograph[:,1::20]}, one processed spatial pixel corresponds to 20 raw kymograph columns. No absolute $\mu$m-per-pixel conversion was applied.

\subsection{Data processing}

To compare waves under slowly changing biochemical conditions, the full time series was divided into eight temporal phases using frame boundaries 120, 180, 240, 300, 360, 420, and 480 (Fig.~S7A). Each phase was treated as a quasi-stationary window for estimating the extract energy-supply state and wave statistics. The final phase was retained as a diagnostic of late-stage behavior but excluded from the correlation analyses in main-text Fig. 4C,D because wave quality deteriorates in the terminal aging regime.

Within each phase, we extracted the wave speed $c$, amplitude $A$, maximal rising slope, and positional fluctuation of the tracked wavefront. Speeds and displacements were computed over lags of 1--12 frames (Fig.~S7C). The displacement variance at each lag was estimated from all tracked positions separated by that lag, and its uncertainty was estimated by 100 bootstrap resamplings. We then obtained a variance-growth coefficient from a weighted linear fit of displacement variance versus frame lag (Fig.~S7D--F). This coefficient is proportional to $D_X$ up to the conventional factor of two in $\mathrm{Var}[X(t)]\simeq 2D_Xt$.

We next linked these image-based quantities to the energetic quantities in the theory. The average maximal rising slope was used as a protein-synthesis-rate proxy because steeper trigger-wave fronts report larger local biochemical fluxes in the aging extract (Fig.~S7B). The FHN trigger-wave estimate predicts $\dot{W}_{\text{prop}}\propto c A m_{\rm h}$ in the recovery-dominated regime. Experimentally, the SiR-tubulin amplitude showed substantial cell-to-cell
variation within each quasi-stationary phase and was positively correlated
with the maximal rising slope in all phases (Fig.~S8A). Moreover, the phase
averages showed the same positive relation between amplitude and maximal
rising slope (Fig.~S8B). This behavior indicates that the wave amplitude is
not fully saturated in the measured regime. We therefore interpret the
system as operating in an activator-unsaturated regime, where the inhibitor
amplitude remains coupled to the activator excursion and can be used as an
empirical proxy for the amplitude factor in \(a_{\rm h} m_{\rm h}\). We then constructed
a propagation-dissipation proxy from the measured speed-amplitude product
and compared it with the independently defined wavefront-activity proxy
across quasi-stationary phases. 

\begin{figure}
    \centering
    \includegraphics[width=0.9\textwidth]{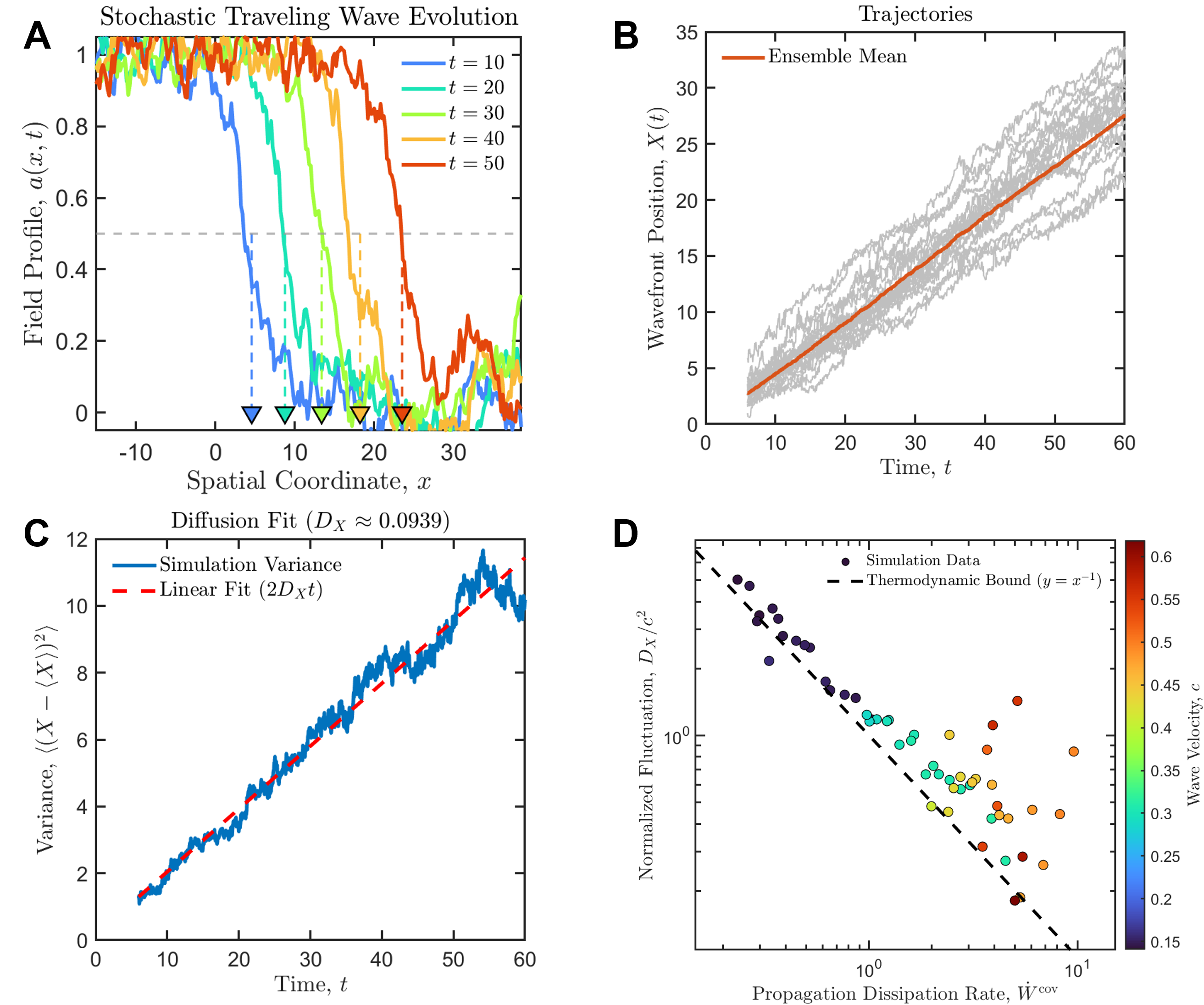} 
    \caption{\textbf{Wave-position diffusion and dissipation in the Allen-Cahn model.}
(A) Representative stochastic wave profiles at different times. The triangles mark $X(t)$, defined as the midpoint between the linearly interpolated $a=0.8$ and $a=0.2$ threshold-crossing positions on the descending front. (B) Stochastic wavefront trajectories (gray curves) and their ensemble mean (orange curve). (C) The variance of the wavefront displacement grows linearly with time, $\mathrm{Var}[X(t)] \simeq 2D_X t$. (D) The normalized fluctuation $D_X/c^2$ decreases with the propagation dissipation rate $\dot{W}^{\text{cov}}$. The dashed line shows the inverse thermodynamic bound, and points are colored by the wave speed $c$.}
\end{figure}

\begin{figure}
    \centering
    \includegraphics[width=0.9\textwidth]{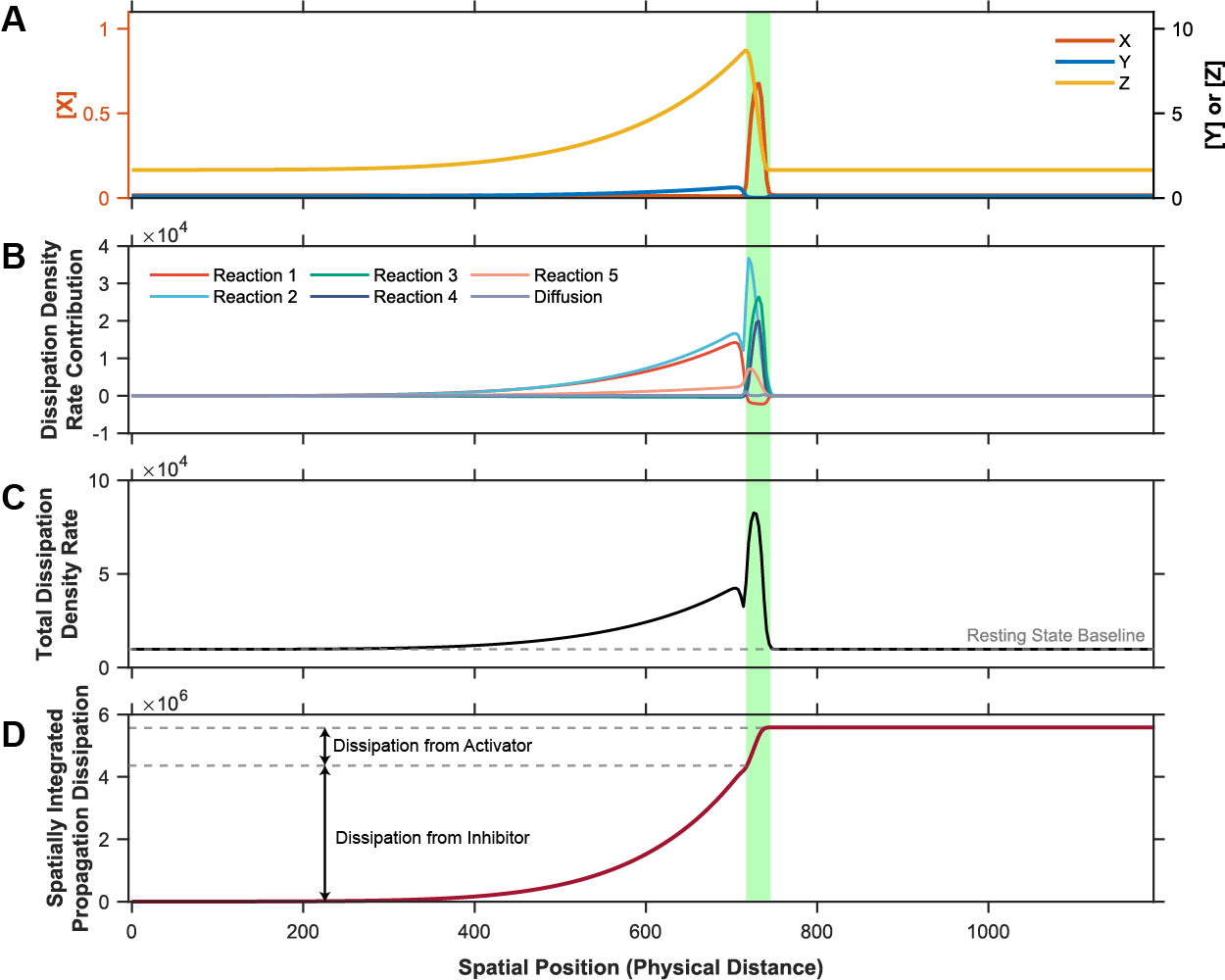} 
    \caption{\textbf{Spatial structure of propagation dissipation in the reversible BZ wave.}
(A) Concentration profiles of the BZ species $X$, $Y$, and $Z$. The shaded green region marks the wavefront region. (B) Local dissipation rate density from individual reactions and diffusion. (C) Overall dissipation rate density, including contributions from all reactions and diffusion. The dashed line denotes the rest-state dissipation rate density. (D) Spatially integrated excess propagation dissipation after subtracting the rest state baseline.}
\end{figure}
\begin{figure}
    \centering
    \includegraphics[width=0.9\textwidth]{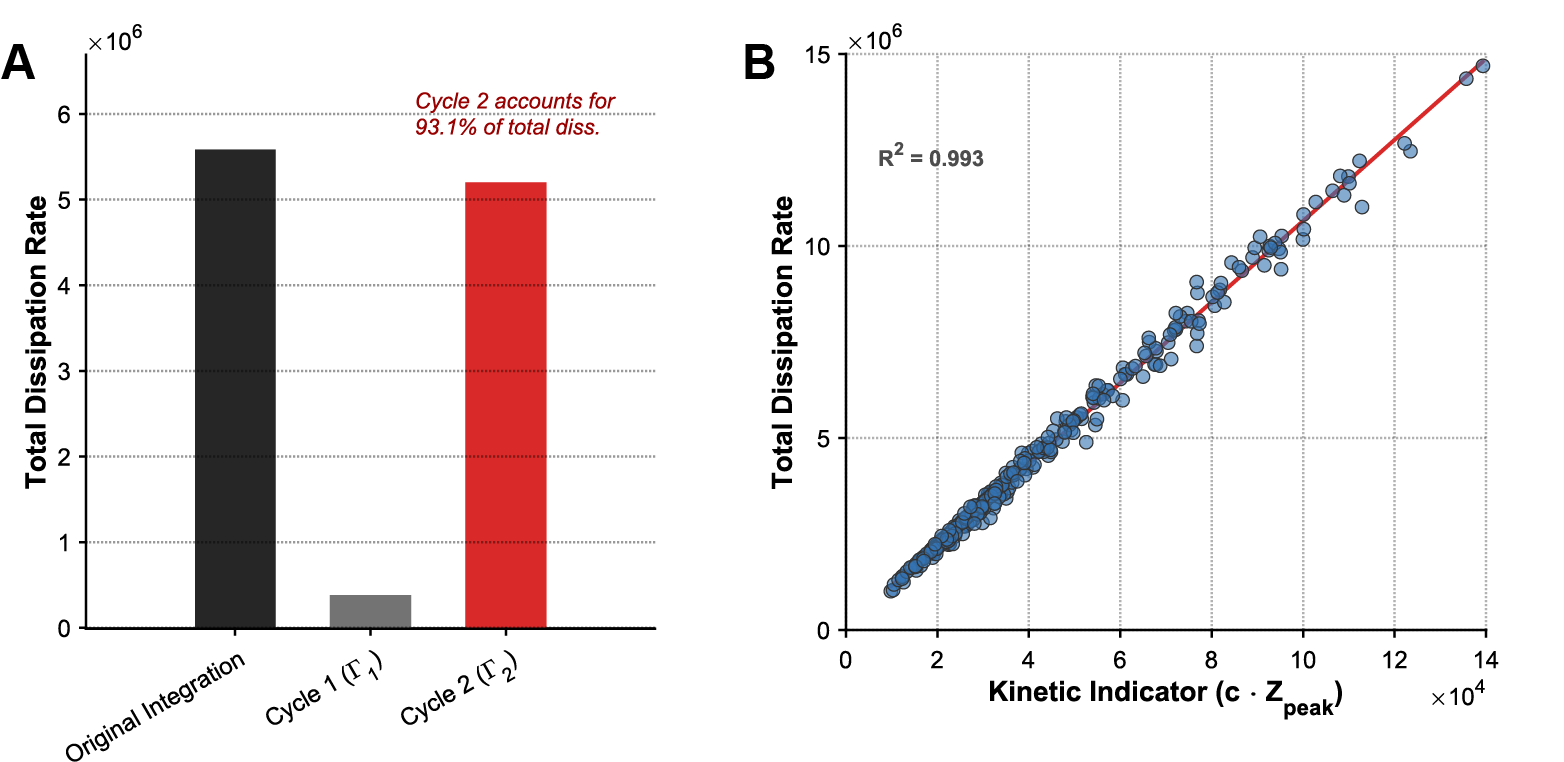} 
    \caption{\textbf{Cycle decomposition and kinetic proxy for BZ propagation dissipation.}
(A) Decomposition of the total dissipation rate into independent thermodynamic cycles in a representative parameter regime. Cycle 2 accounts for most of the dissipation. (B) The total dissipation rate is approximately proportional to the product of the wave speed $c$ and the excess inhibitor peak $Z_{\mathrm{peak}}$, supporting $cZ_{\mathrm{peak}}$ as a kinetic proxy for propagation dissipation.}
\end{figure}
\begin{figure}
    \centering
    \includegraphics[width=0.9\textwidth]{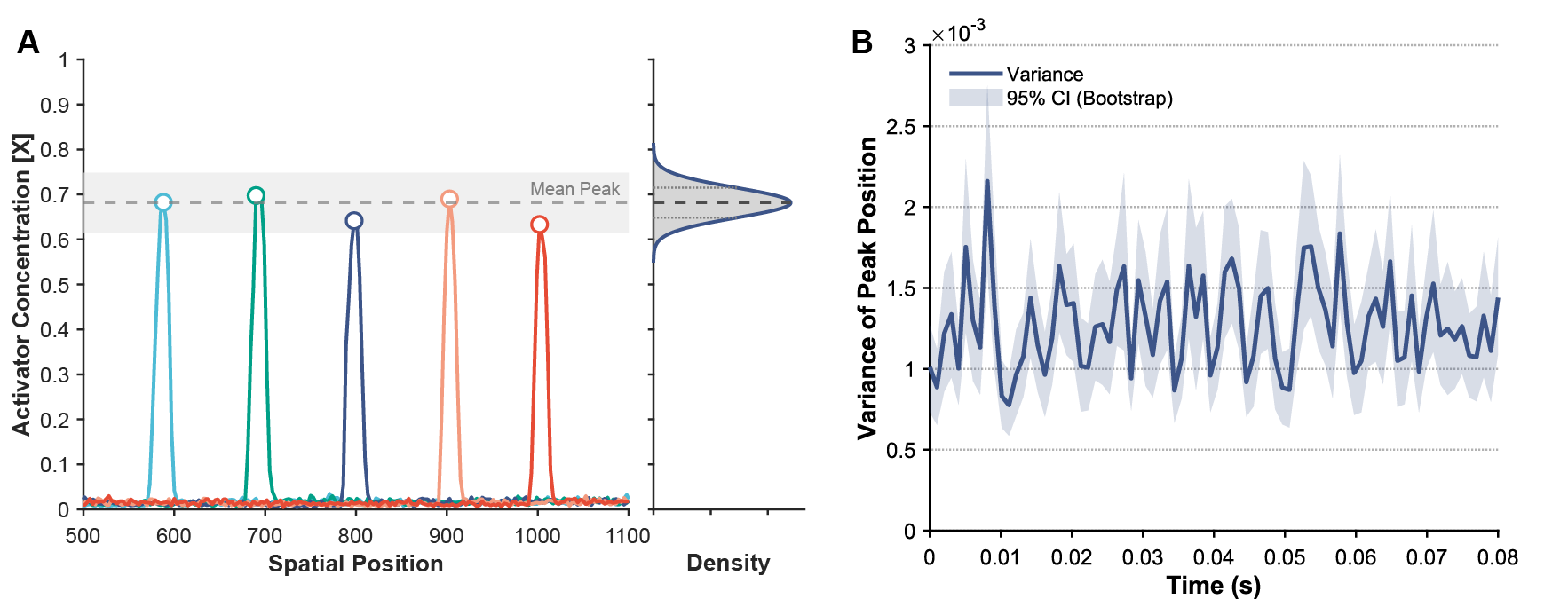} 
    \caption{\textbf{Bounded shape fluctuations.}
(A) Representative activator profiles during propagation. The peak height fluctuates around a stable mean value; the gray band and side probability density show the typical amplitude distribution. (B) The variance of the activator peak height remains bounded over time.}
\end{figure}
\begin{figure}
    \centering
    \includegraphics[width=0.9\textwidth]{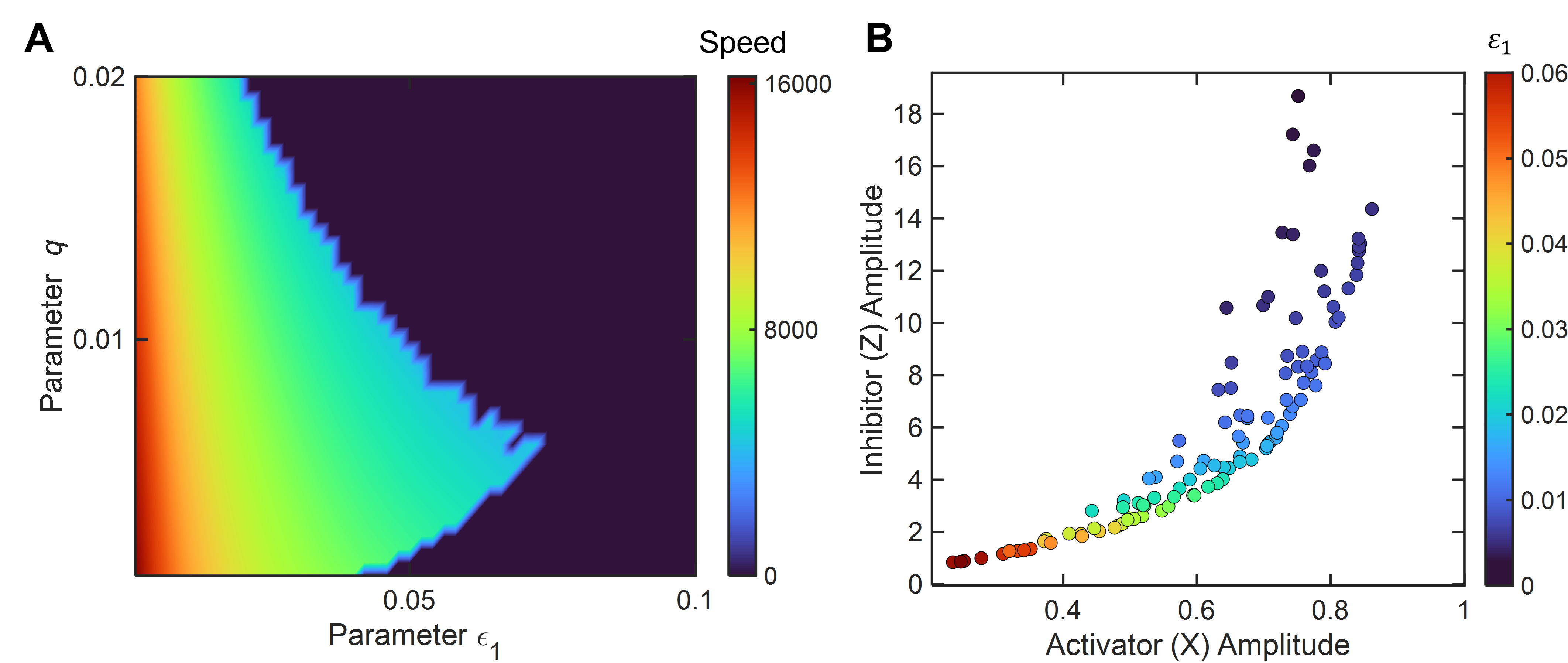} 
    \caption{\textbf{Deterministic parameter scan in the BZ model.}
(A) Wave speed over the $(\epsilon_1,q)$ parameter plane of the reduced BZ model. The colored region denotes stable propagating waves, whereas the dark region corresponds to parameter sets without a stable wave solution. (B) Relationship between activator $X$ amplitude and inhibitor $Z$ amplitude across the propagating-wave regime, colored by $\epsilon_1$.}
\end{figure}
\begin{figure}
    \centering
    \includegraphics[width=0.9\textwidth]{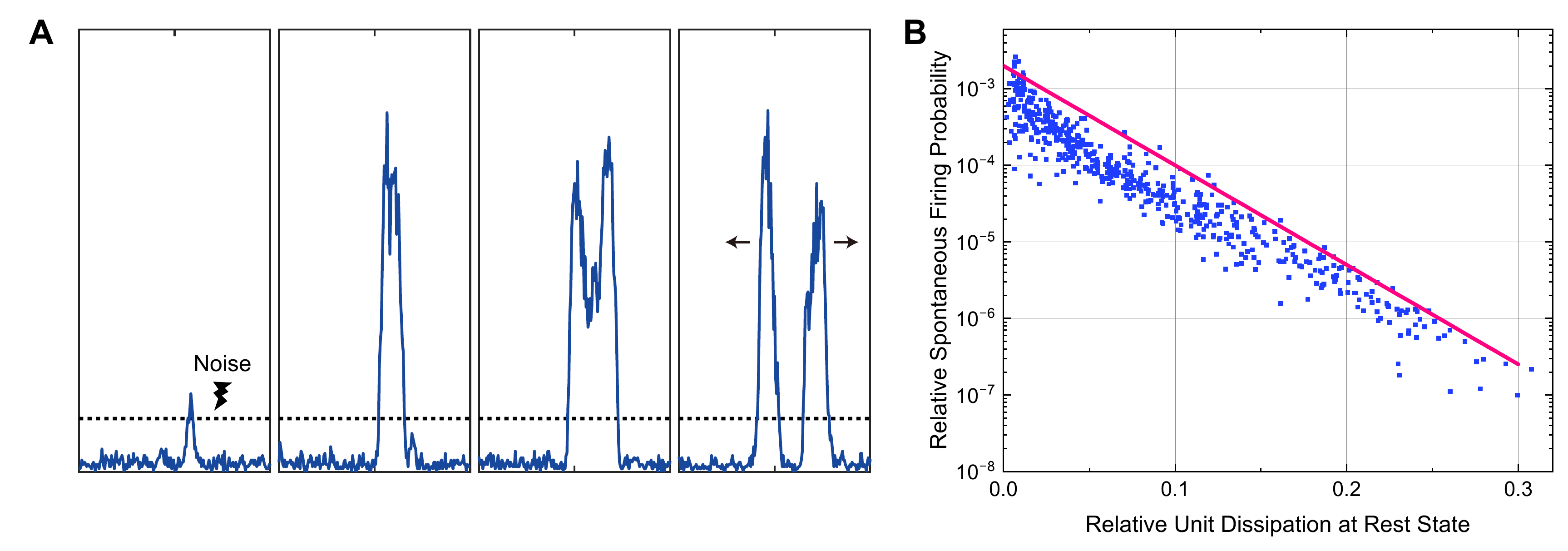} 
    \caption{\textbf{Rest state dissipation suppresses spontaneous firing.}
(A) Representative noisy BZ profiles illustrating spontaneous firing events generated by fluctuations around the rest state. Threshold-crossing fluctuations can nucleate pulses that propagate in opposite directions. (B) The relative spontaneous firing probability decreases as the relative unit dissipation of the rest state increases.}
\end{figure}
\begin{figure}
    \centering
    \includegraphics[width=0.9\textwidth]{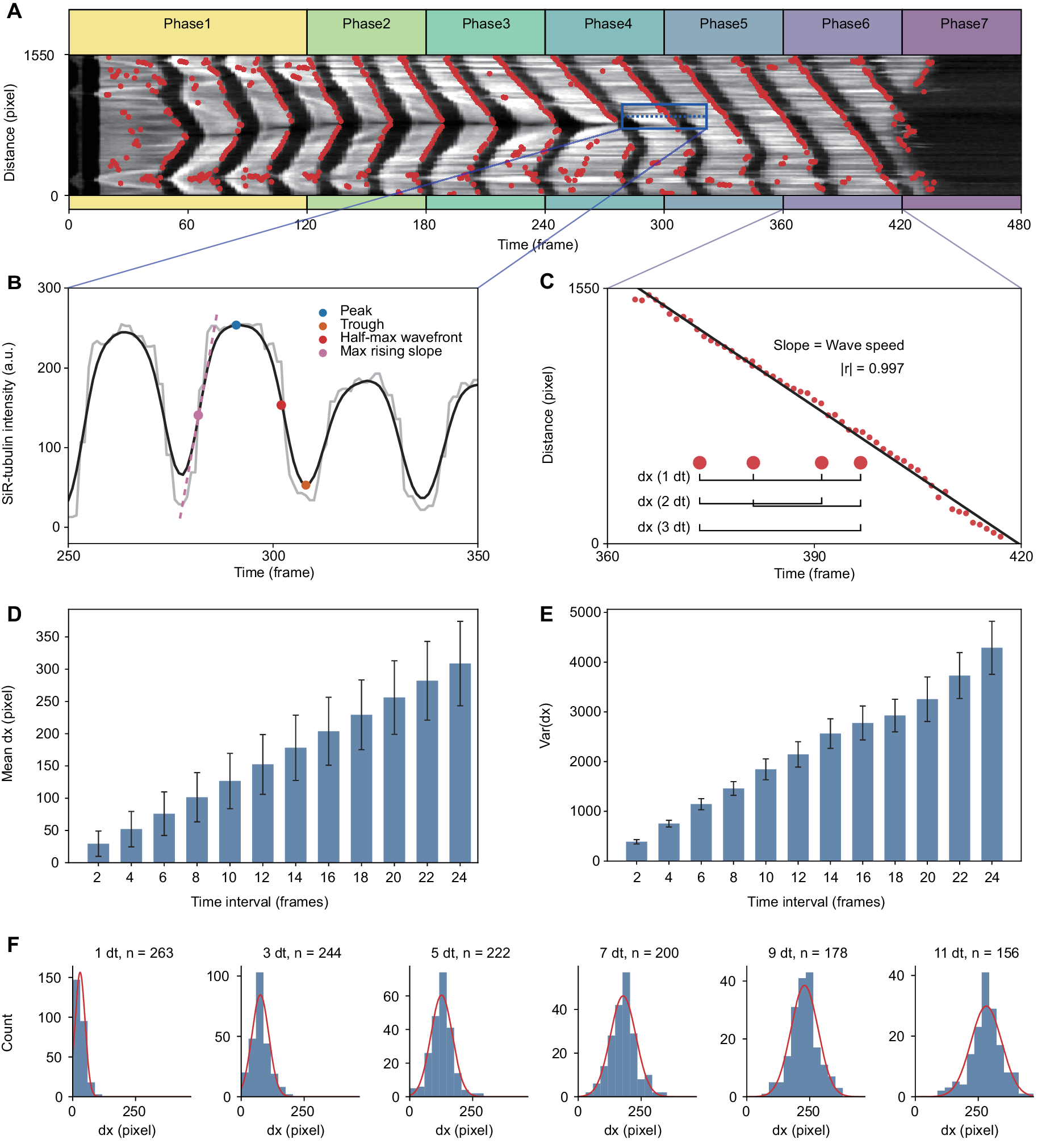} 
    \caption{\textbf{Tracking and fluctuation analysis of mitotic trigger waves in \textit{Xenopus} egg extract.}
(A) Representative SiR-tubulin kymograph with tracked half-maximum wavefront positions as red dots. Colored bars indicate the quasi-stationary temporal phases used to compare wave statistics as the extract ages. (B) Local time trace used to identify peaks, troughs, half-maximum wavefront positions, and the maximal rising slope. (C) Linear fit to a tracked wavefront segment. The slope gives the wave speed. Displacements $dx$ over multiple frame lags are used to estimate wave wandering. (D,E) Mean displacement and displacement variance as functions of time interval. (F) Displacement distributions for representative time lags, with fitted distributions used to summarize the fluctuation statistics.}
\end{figure}
\begin{figure}
    \centering
    \includegraphics[width=0.9\textwidth]{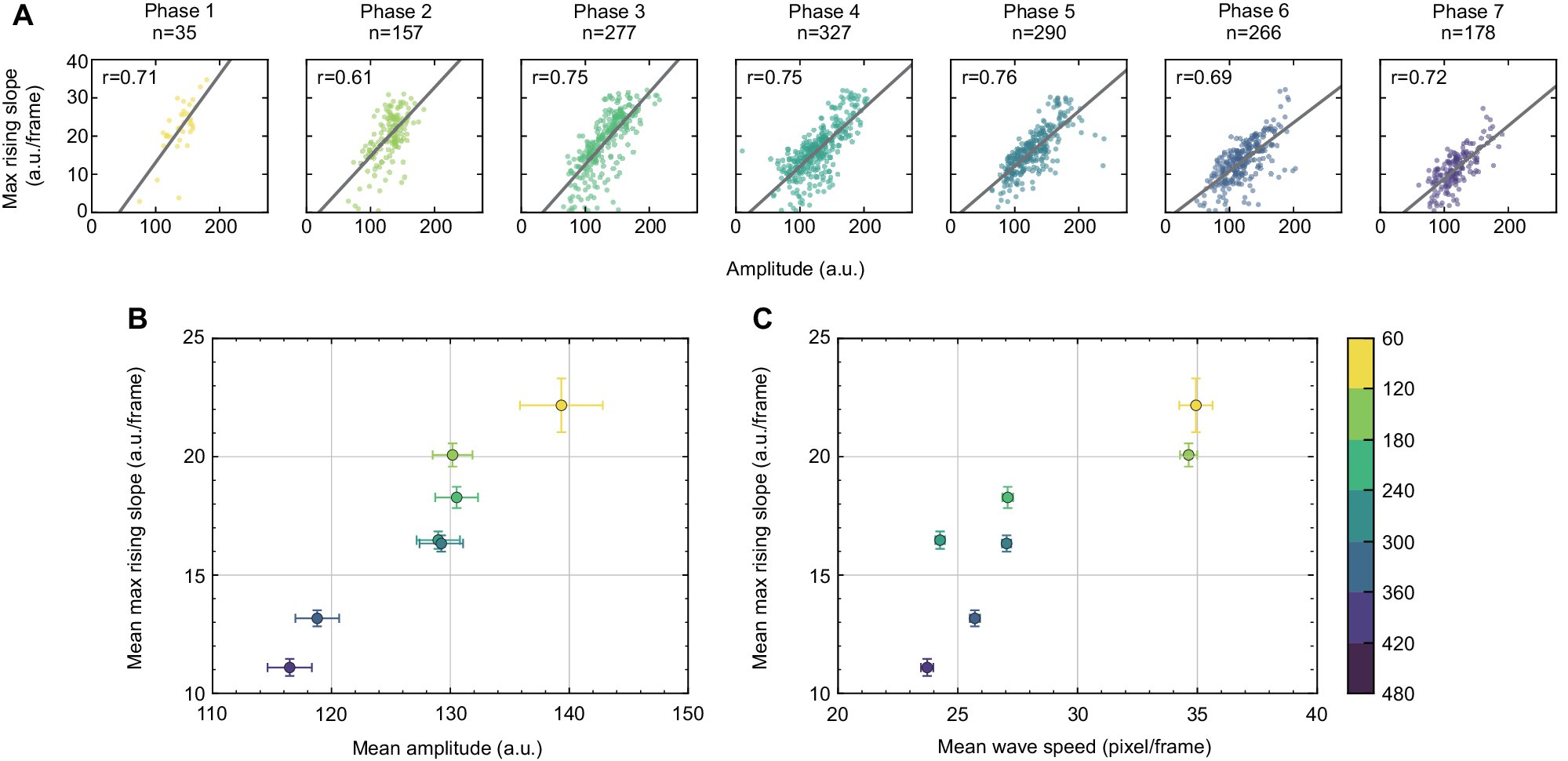} 
    \caption{\textbf{Correlations of the maximal rising slope with wave amplitude and wave speed.}
(A) Within each quasi-stationary phase, the maximal rising slope of the SiR-tubulin signal is positively correlated with the wave amplitude. Points denote individual wavefront measurements, gray lines show linear fits, and $r$ is the Pearson correlation coefficient. (B) Phase-averaged maximal rising slope versus phase-averaged amplitude. (C) Phase-averaged maximal rising slope versus phase-averaged wave speed. Colors indicate the temporal phase of the aging extract. }
\end{figure}